\documentclass{jpp}
\usepackage{graphicx}

\usepackage[utf8]{inputenc}
\usepackage[T1]{fontenc}
\usepackage{amsmath,amssymb,bm,xcolor}

\newcommand{\Alfven}{Alfv\'{e}n}
\newcommand{\Alfvenic}{Alfv\'{e}nic}

\newcommand{\V}[1]{\mathbf{#1}}

\newcommand{\xhat}{\mbox{$\hat{\mathbf{x}}$}}
\newcommand{\yhat}{\mbox{$\hat{\mathbf{y}}$}}

\newcommand{\figref}[1]{Fig.~\ref{#1}}   
\newcommand{\eqr}[1]{Eq.~\eqref{#1}}   
\newcommand{\secref}[1]{\S\ref{#1}}
  
\newcommand{\pfrac}[2]{\frac{\partial #1}{\partial #2}}

\newcommand{\grad}{\mathbf{\nabla}}
\newcommand{\Vu}{\mathbf{u}}
\newcommand{\Vv}{\mathbf{v}}
\newcommand{\VB}{\mathbf{B}}
\newcommand{\Vz}{\mathbf{z}}
\newcommand{\Vk}{\mathbf{k}}
\newcommand{\hx}{\hat{\mathbf{x}}}
\newcommand{\hy}{\hat{\mathbf{y}}}
\newcommand{\hz}{\hat{\mathbf{z}}}
\newcommand{\order}[1]{\mathcal{O}\left(#1\right)}
\newcommand{\BHAC}{{\tt BHAC}}

\newcommand{\alert}[1]{\textcolor{red}{#1}}

\shorttitle{Weak relativistic Alfv\'{e}nic turbulence}
\shortauthor{J. M. TenBarge, et al.}

\title{Weak Alfv\'{e}nic turbulence in relativistic plasmas. Part 1. Dynamical equations and basic dynamics of interacting resonant triads}

\author{J.~M. TenBarge\aff{1,2}\corresp{\email{tenbarge@princeton.edu}},
  B. Ripperda\aff{1,2},
  A. Chernoglazov\aff{2,3},
  A. Bhattacharjee\aff{1,2},
  J.~F. Mahlmann\aff{1},
  E.~R. Most\aff{4,5,6},
  J. Juno\aff{7},
  Y. Yuan\aff{2},
  \and  A.~A. Philippov\aff{2}}
  
\affiliation{
\aff{1}Department of Astrophysical Sciences, Peyton Hall, Princeton University, Princeton, NJ 08544, USA
\aff{2}Princeton Center for Heliophysics, Princeton University, Princeton, NJ 08540
\aff{3}Center for Computational Astrophysics, Flatiron Institute, 162 Fifth Avenue, New York, NY 10010, USA
\aff{4}Department of Physics, University of New Hampshire, 9 Library Way, Durham NH 03824, USA
\aff{5}Princeton Center for Theoretical Science, Jadwin Hall, Princeton University, Princeton, NJ 08544, USA
\aff{6}Princeton Gravity Initiative, Princeton University, Princeton, NJ 08544, USA
\aff{7}School of Natural Sciences, Institute for Advanced Study, Princeton, NJ 08544, USA
\aff{8}Department of Physics and Astronomy,University of Iowa, Iowa City IA 52242, USA}

\begin{document}

\maketitle

\begin{abstract}
\Alfven\ wave collisions are the primary building blocks of the non-relativistic turbulence that permeates the heliosphere and low-to-moderate energy astrophysical systems. However, many astrophysical systems such as gamma-ray bursts, pulsar and magnetar magnetospheres, and active galactic nuclei have relativistic flows or energy densities. To better understand these high energy systems, we derive reduced relativistic MHD equations and employ them to examine weak \Alfvenic\ turbulence, {dominated by three-wave interactions,} in reduced relativistic magnetohydrodynamics, including the force-free, infinitely magnetized limit. We compare both numerical and analytical solutions to demonstrate that many of the findings from non-relativistic weak turbulence are retained in the relativistic system. But, an important distinction in the relativistic limit is {the {inapplicability of a} formally incompressible limit, i.e, there exists finite coupling to the compressible fast mode regardless of the strength of the magnetic field.} Since fast modes can propagate across field lines, this mechanism provides a  route for energy to escape strongly magnetized systems, e.g., magnetar magnetospheres. However, we find that the fast-\Alfven\ coupling is diminished in the limit of oblique propagation. 

\end{abstract}

\section{Introduction}\label{sec:intro}
Turbulence provides the transport of mass, momentum, and energy in a wide range of plasmas throughout the universe, from the intracluster medium and magnetar magnetospheres to the solar wind and laboratory fusion confinement experiments. In space and astrophysical plasmas, turbulence plays the important role of transferring large scale motions, often driven by violent processes and instabilities, to small scales at which damping, dissipation, and plasma heating can occur. This cascade of turbulent energy is governed by nonlinear interactions that occur within the plasma, and the process is well studied in the non-relativistic (Newtonian) limit. However, many astrophysical plasmas of interest are relativistic and often magnetically dominated ($b^2 \gg h$, where $b^2 = b_\mu b^\mu = B^2/\gamma^2 + \left(\bm{B}\cdot\bm{v}\right)^2$ is the magnetic energy density, $h$ is the enthalpy density, and $\gamma$ is the Lorentz factor). How these energetic plasmas are heated is fundamental for interpreting the electromagnetic radiation we observe at Earth.

We will begin our study of relativistic, magnetically dominated turbulence by reviewing and building upon results from Newtonian turbulence. The magnetic fields that universally permeate plasmas imply that \Alfvenic\ fluctuations \citep{Alfven:1942} will govern the hierarchy of turbulent fluctuations rather than the eddies that compose hydrodynamic turbulence. The shear (or transverse) \Alfven\ wave has the property that the fluid motions corresponding to it are entirely transverse to the background magnetic field,  with no compressional component. Based on these ideas, \cite{Iroshnikov:1963} and \cite{Kraichnan:1965} (IK) employ incompressible magnetohydrodynamics (MHD) to propose that the nonlinear interactions in plasma turbulence are composed of counter-propagating and overlapping \Alfven\ waves or wave packets. 

The fundamental assumption underlying the IK picture of turbulence is that the so-called $\V{E} \times \V{B}$ nonlinearity is the dominant nonlinear term in the plasma. In terms of fluid, MHD, equations, this term appears as $\V{v}_\perp \cdot \nabla \star$, where $\V{v}_\perp \simeq \V{E} \times \V{B}_0/B_0^2$ is the dominant drift velocity in the plasma, $\V{B}_0$ is a mean magnetic field, $\star$ indicates either $\Vv$ or $\VB$, and perpendicular, $\perp$, indicates perpendicular to $\V{B}_0$. The form of this term makes clear immediately that the nonlinearity requires fluctuations in the plane perpendicular to the mean magnetic field, i.e., $k_\perp \ne 0$. This term also requires that the {perpendicular wave vectors of} interacting \Alfven\ waves {are not collinear}, but this point is not obvious from the simplified expression above and will be explored further in  \secref{sec:incompMHD}. Although there are cases in which the $\V{E} \times \V{B}$ nonlinearity is not dominant, e.g., parametric instabilities such as the decay, modulation, and beat instabilities, we  will  adopt the assumption that the  $\V{E} \times \V{B}$ nonlinearity is the dominant nonlinear term. Additionally, the above discussion is based on a fluid description of plasma; however, most high energy astrophysical systems are in the weakly collisional limit ($l \ll \lambda_{mfp}$, where $l$ is an intermediate turbulence scale and $\lambda_{mfp}$ is the collisional mean free path), formally requiring a kinetic description. Fortunately, kinetic Newtonian turbulence retains many of the same basic properties as MHD turbulence for scales larger than ion kinetic scales, e.g., the ion inertial length and gyroradius. Importantly, the $\V{E} \times \V{B}$ nonlinearity is the dominant nonlinearity in kinetic Newtonian turbulence, even at turbulence scales below ion kinetic scales \citep{Schekochihin:2009}.

The IK theory of \Alfvenic\ turbulence assumes that the turbulent cascade proceeds isotropically across scales, i.e.,  $k_\parallel \sim k_\perp$ at all scales. \cite{Montgomery:1981} and \cite{Shebalin:1983} revisit the IK theory under the assumption that the turbulence is weak, $\omega_{NL} \ll \omega$, where $\omega_{NL} \sim k_\perp \delta v_\perp$ is the nonlinear frequency and $\omega \sim k_\parallel v_A$ is the linear, \Alfven\ frequency. Under this weak assumption, they demonstrate that the three-wave interaction is the dominant nonlinear interaction of the $\V{E} \times \V{B}$ nonlinearity. Importantly, this three-wave interaction involves a frequency and wavevector matching condition for two counter-propagating \Alfven\ waves colliding to produce a $k_\parallel = 0$ mode, leading to an anisotropic cascade of energy which fixes $k_\parallel$ at the outer-scale. Based on this interaction, \cite{Montgomery:1981,Montgomery:1982,Higdon:1984a} develop a theory of highly anisotropic incompressible MHD turbulence consisting of two-dimensional velocity and magnetic field fluctuations in the plane perpendicular to the mean magnetic field\footnote{Note that in the purely 2D limit, i.e., the 2D plane not containing the mean field, the distinction between weak and strong \Alfvenic\ turbulence vanishes, since the linear physics of \Alfven\ waves is mostly eliminated. Therefore, 2D simulations in this geometry are always strong, analogous to hydrodynamic turbulence that does not support linear waves.}. To describe this system, they employ the reduced MHD (RMHD) equations introduced by \cite{Kadomstev:1974,Strauss:1976} for systems in which there is a strong mean magnetic field, leading to the ordering assumptions of RMHD: $\delta B_\perp / B_0 \sim \delta v_\perp / v_A \sim k_\parallel / k_\perp \sim \epsilon \ll 1$. Note that the assumptions of RMHD necessarily imply that the \Alfven\ and fast modes are well separated in frequency since $\omega_{A} \propto k_\parallel \ll \omega_F \propto k \sim k_\perp$. This ordering implies that the fast mode is ordered out of the RMHD system. All of the following discussion regarding the development of MHD turbulence is predicated on this assumption that the fast wave is well separated from the \Alfven\ mode and is therefore ignorable. To develop a theory of relativistic, magnetically dominated astrophysical plasmas, we will revisit the concept of reduced MHD in the relativistic limit in section \secref{sec:RRMHD}.

\cite{Sridhar:1994} take a different approach in expanding upon the work of IK by first pointing out that IK is a theory of weak turbulence, and then loosening the isotropy assumption of IK by building anisotropy into weak turbulence theory. However, \cite{Sridhar:1994} argue that the three-wave interaction is empty because it involves a mode with $\omega = 0$, which cannot be a linear wave mode with finite power. Therefore, they invoke a four-wave interaction ($\mathrm{A}+\mathrm{A} \rightarrow \mathrm{A} + \mathrm{A}$) which maintains $k_\parallel = 0$, and as the weak cascade proceeds to smaller scales, the strength of nonlinear interactions increases, eventually reaching a state of strong turbulence. \cite{Goldreich:1995} carry the weak turbulence theory developed in \cite{Sridhar:1994} into the strong limit. In the strong limit, they argue that the resonance condition for interaction is broadened, and due to the broadening, a parallel cascade develops. Further, they argue that the parallel cascade leads the turbulence towards a state of critical balance in which $\chi = \omega_{NL} / \omega \simeq 1$, where $\chi$ is the nonlinearity parameter and $\chi \ll 1$ corresponds to weak turbulence. Critical balance is a condition in which the nonlinear frequency or cascade time is balanced by the linear time in the system. The critically balanced turbulence cascade is predicted to have an energy spectrum $E \sim k_\perp^{-5/3}$ and a spectral anisotropy of the form $k_\parallel \sim k_\perp^{2/3}$, thus, developing a scale dependent anisotropy. Note that \Alfvenic\ turbulence, whether weak or strong, always leads to a case at small scales in which $k_\parallel \ll k_\perp$ and $\delta B \ll B_0$ regardless of the isotropy or amplitude of fluctuations at the outer-scale\footnote{Built into this statement and all discussions of turbulence herein is the assumption that the viscosity and resistivity are sufficiently small that a turbulent cascade is able to fully develop.}. 

\cite{Montgomery:1995} and \cite{Ng:1996} concurrently claim that \cite{Sridhar:1994} are incorrect to claim that three-wave interactions are empty, because $k_\parallel = 0$ modes are valid nonlinear fluctuations. \cite{Ng:1997} further employ perturbation theory to demonstrate explicitly that the three-wave interaction is non-empty, $k_\parallel = 0$ modes do develop, and the three-wave interaction dominates over the four-wave. They also present the energy spectrum of weak turbulence, $E \sim k_\perp^{-2}$. Admitting their mistake in omitting the three-wave interaction, \cite{Goldreich:1997} reformulate their weak turbulence theory to include three-wave interactions. The basic prediction for the spectral anisotropy (no parallel cascade), energy spectrum $E \sim k_\perp^{-2}$, and the strengthening of the cascade as it proceeds to smaller scales remain unchanged. \cite{Galtier:2000} provide a rigorous derivation of the $k_\perp^{-2}$ scaling result based on the wave-kinetic approach \citep{Zakharov:1992}{, and \cite{Perez:2008} numerically verify the derivation and examine the weak to strong turbulence transition}. \cite{Lithwick:2003} present further extensions of weak turbulence theory.

More recently, a series of papers were written examining {the building blocks of} weak turbulence {through the interaction of waves} analytically \citep{Howes:2013a}, numerically \citep{Nielson:2013}, and in an experiment \citep{Howes:2013b,Drake:2013} conducted at the Large Plasma Device (LaPD) \citep{Gekelman:1991}. This series of papers focuses on {a heuristic, analytical} solution beginning with a collision of counter-propagating \Alfven\ waves at first order {in the fluctuation amplitude} and following their evolution order-by-order. At second order, the three-wave interaction involving the nonlinear, $k_\parallel = 0$ mode is found to be dominant. This mode does not involve a secular exchange of energy and oscillates with frequency $\omega = 2\omega_0$, where $\omega_0 = k_\parallel v_A$ is the frequency of the incident \Alfven\ waves. At third order, the incident \Alfven\ waves interact with the $k_\parallel = 0$ mode, and the $k_\parallel = 0$ mode shears the \Alfven\ waves in the perpendicular plane, providing a secular exchange of energy to smaller perpendicular scales but fixed $k_\parallel$ scale.
 
 The theory of critically balanced strong turbulence has also been further refined since \cite{Goldreich:1995}. \cite{Boldyrev:2005,Boldyrev:2006} note that the vector nature of the nonlinearity leads to a state called dynamic alignment wherein the velocity and magnetic field fluctuations align themselves as the cascade proceeds to smaller scales. This alignment leads to the formation of thin current sheets at small scales, and these current sheets can eventually disrupt \citep{Mallet:2017,Loureiro:2017,Comisso:2018,Dong:2018}. Intermittency has also been built into the theory of critically balanced and dynamically aligned turbulence \citep{Chandran:2015}, leading to the theory of refined critical balance \citep{Mallet:2015}. For a more complete, contemporary (but biased, according to the authors) review of the current state of MHD turbulence including weak, strong, and imbalanced turbulence, see \cite{Schekochihin:2020}.
 
 Although fundamentally important for understanding energy dissipation and astrophysical observations, relativistic turbulence has received much less attention. \cite{Thompson:1998,Troischt:2004} examine weak turbulence in the magnetically dominated, relativistic regime. \cite{Thompson:1998} follow \cite{Sridhar:1994} and \cite{Goldreich:1995} in arguing that the four-wave interaction is the dominant \Alfvenic\ interaction because the $k_\parallel = 0$ mode is not a linear mode of the system. However, they note that in the extreme relativistic regime, one cannot assume that the fast and \Alfven\ modes are well separated, i.e., the intuition gained from incompressible MHD no longer holds. Therefore, they argue that the dominant three-wave interactions are of the form $\mathrm{A}+\mathrm{A} \rightarrow \mathrm{F}$, $\mathrm{A}+\mathrm{F} \rightarrow \mathrm{A}$, or $\mathrm{F}+\mathrm{F} \rightarrow \mathrm{F}$. \cite{Heyl:1999} extend the work of \cite{Thompson:1998} to include {quantum electrodynamic} effects making the same assumptions regarding three versus four-wave interactions. \cite{Li:2019b} return to the problem of weak, magnetically dominated turbulence following the work of \cite{Thompson:1998}, again assuming that the dominant \Alfvenic\ interactions are the four-wave coupling or the three-wave couplings involving fast modes, as listed above. \cite{Li:2019b} follow the analytical discussion with relativistic MHD simulations in the force-free limit considering both weak and strong turbulence limits. They focus on the nonlinear, turbulent conversion of \Alfven\ to fast mode energy as a possible mechanism to release energy from magnetar magnetospheres, since \Alfven\ waves propagate along fields lines and remain trapped in the magnetosphere, while fast modes can propagate across fields lines, thereby escaping confinement and releasing energy. A variety of other recent papers have explored various aspects of relativistic turbulence theory and simulations, finding broadly similar results to Newtonian MHD turbulence, except they consistently find a small portion ($\lesssim 10-15\%$) of the initial \Alfvenic\ energy leaks into the fast mode branch \citep{Cho:2005,Takamoto:2016,Takamoto:2017,Li:2019b}.
 
 This work is organized as a sequence of papers. In this manuscript (Paper I), {we derive a set of relativistic reduced MHD (RRMHD) equations which have the same form and properties as their Newtonian counterpart, including the wave kinetic equation governing spectral evolution.} We then employ an approach similar to \cite{Howes:2013a} to obtain an {analytical} solution for {three-wave interactions} in relativistic systems. {We emphasize that although we make a connection with the wave kinetic equation for the relativistic system, the primary analysis is intended to be heuristic to highlight the role of the three-wave interactions and other similarities with Newtonian turbulence rather than a formal weak turbulence theory.} {Our approach, wherein we obtain the solution order-by-order,  is well suited to comparison with numerical simulations (Paper II) and complements the variational approach of \cite{Thompson:1998}.} We build upon the wisdom gained from Newtonian turbulence and begin by outlining {and reviewing} some of the salient properties {for an astrophysical audience} of both incompressible and compressible MHD turbulence in \secref{sec:general}. In \secref{sec:equations}, we derive the relativistic Elsasser equations in the reduced, relativistic limit {and discuss their connection with weak Newtonian turbulence}. We then derive through third order the weak \Alfvenic\ turbulence solutions in \secref{sec:asymptotic}. In \secref{sec:simulations}, we compare our solutions to direct numerical simulations and consider the role of fast waves in both relativistically hot and magnetically dominated turbulence. Finally, we summarize the results in \secref{sec:conclusions}. More in-depth numerical simulation studies of weak, relativistic turbulence are presented in the second paper of the sequence \citep{Ripperda:2021}, henceforth Paper II.

\section{General Properties of Non-Relativistic Turbulence}\label{sec:general}
\subsection{Incompressible MHD}\label{sec:incompMHD}

Before exploring relativistic \Alfvenic\ turbulence, it is important to first review some of the fundamental knowledge learned from Newtonian turbulence to provide a framework for discussing the relativistic limit. This discussion is far from exhaustive, because incompressible MHD turbulence is an exceptionally broad and deep field. 

We will begin our discussion by presenting some of the basic properties of incompressible turbulence. Incompressible MHD is the basis from which Newtonian turbulence theories have been derived. Incompressible MHD assumes $v \ll c_s$, where $c_s$  is the sound speed. In other words, compressive fluctuations are carried away from the source at essentially infinite speed, an assumption that is not applicable for relativistic systems. This assumption implies $\grad \cdot \V{v} = 0$, leading to the incompressible MHD equations
\begin{equation}\label{eq:incMom}
\pfrac{\Vv}{t} + \Vv \cdot \grad \Vv = -\grad P/\rho_0 + \VB \cdot \grad \VB/\rho_0,
\end{equation}
\begin{equation}\label{eq:incInd}
    \pfrac{\VB}{t} + \Vv \cdot \grad \VB = \VB \cdot \grad \Vv,
\end{equation}
\begin{equation}
    \grad \cdot \VB = 0,
\end{equation}
where $P$ is the total (thermal plus magnetic) pressure, and $\rho_0$ is the equilibrium mass density.

For turbulence analysis, the incompressible MHD equations are typically cast in Elsasser form \citep{Elsasser:1950} by adding and subtracting the momentum and induction equations to arrive at
\begin{equation}\label{eq:NRElsasser}
    \pfrac{\Vz^\pm}{t} \mp \V{v}_A \cdot \grad \Vz^\pm = -\Vz^\mp \cdot \grad \Vz^\pm -\grad P / \rho_0,
\end{equation}
\begin{equation}
    \grad \cdot \Vz^\pm = 0,
\end{equation}
    where the magnetic field has been separated into equilibrium and fluctuating parts, $\VB = B_0 \hat{\Vz} + \delta \VB$, $\V{v}_A = \VB_0 / \sqrt{4 \pi \rho_0}$,  and $\Vz^\pm = \Vv \pm \delta \VB /  \sqrt{4 \pi \rho_0}$ are the Elsasser fields. Because the Elsasser fields are divergenceless, this is a closed system of equations \citep{Montgomery:1982}. To obtain an equation for the pressure, one can simply take the divergence of \eqr{eq:NRElsasser} to find
\begin{equation}
    \nabla^2 P/\rho_0 = -\grad \cdot \left( \Vz^\mp \cdot \grad \Vz^\pm\right).
\end{equation}

The Elsasser equations, \eqr{eq:NRElsasser}, are the progenitor equations to describe \Alfvenic\ turbulence, and one can discern many important points about turbulence simply from the form of the equations. First, the terms on the left hand side are linear, while those on the right are nonlinear. Therefore, linearizing the equations is as simple as setting the right hand side to zero. By linearizing, one can immediately see that the system supports two propagating linear wave modes with $\omega = \pm k_\parallel v_A$: (i) \Alfven\ waves with $\Vz^\pm$ polarized in the $\hat{\Vz} \times \hat{\V{k}}^\pm$ direction, and (ii)
pseudo-\Alfven\ waves, the incompressible limit of magnetosonic slow modes, with polarization $\hat{\V{k}}^\pm \times \left(\hat{\Vz} \times \hat{\V{k}}^\pm\right)$, where $\hat{\V{k}}^\pm$ corresponds to the unit vector along the wavevector of $\Vz^\pm$. We adopt the convention that $\omega \ge 0$, and the sign of $k_\parallel$ determines the propagation direction. Note that the incompressible assumption orders the magnetosonic fast mode out of the system by {imposing} $c_s \rightarrow \infty$. Thus, one can interpret the Elsasser field $\Vz^+$ ($\Vz^-$) as describing the evolution of \Alfven\ or pseudo-\Alfven\ waves propagating down (up) the equilibrium magnetic field.

Next, we turn our focus to the primary nonlinear term, $\Vz^\mp \cdot \grad \Vz^\pm$. The most immediate point one can see from the form of this term is that the nonlinearity only survives if both $\Vz^+$ and $\Vz^-$ are nonzero, {i.e., the nonlinearity is one that does not involve self-interaction of an \Alfven\ wave with itself but its interaction with an oppositely propagating \Alfven\ wave.} If either Elsasser field is zero, then the opposite Elsasser variable is an exact solution. For instance, if $\Vz^- = 0$, then $\Vz^+(x,y,z+v_A t)$ is an exact, nonlinear solution representing an arbitrary amplitude \Alfven\ or pseudo-\Alfven\ wave propagating in the $-\hat{\Vz}$ direction. 

Physically, the counter-propagating waves shear one another when they interact through this nonlinear term, and the shearing leads to a transfer of energy to smaller scales. For simplicity, let us focus on the case of counter-propagating \Alfven\ waves and examine the nonlinear term in more detail. In Fourier space, the nonlinear interaction term for a $\Vz^+$ wave distorted by a $\Vz^-$ wave is 
\begin{equation}\label{eq:polarization}
\Vz^- \cdot \grad \Vz^+ \propto \left[\left(\hat{\Vz} \times \hat{\V{k}}^- \right) \cdot \hat{\V{k}}^+\right] \left(\hat{\Vz} \times \hat{\V{k}}^+ \right) = \left[\hat{\Vz} \cdot \left( \hat{\V{k}}^- \times \hat{\V{k}}^+ \right)\right] \left(\hat{\Vz} \times \hat{\V{k}}^+ \right),
\end{equation}
where we have used the polarization properties of the \Alfven\ waves to write $\Vz^\pm \propto  \hat{\Vz} \times \hat{\V{k}}^\pm$. Therefore, for the nonlinear interaction of counter-propagating \Alfven\ waves to be nonzero, $\hat{\Vz} \cdot \left( \hat{\V{k}}^- \times \hat{\V{k}}^+ \right) \ne 0$, i.e., {variation is required in both directions perpendicular to the equilibrium field. An alternative way to state this requirement is that} the waves must be polarized with respect to one another in the plane perpendicular to the equilibrium field {so that the perpendicular wave vector components are not collinear.}

From examination of the linear and nonlinear terms of the Elsasser equation, we have gleaned three crucial facts for turbulence: (i) The system supports two linear waves modes, both of which require $k_\parallel \ne 0$ to propagate; (ii) The nonlinearity requires counter-propagating fluctuations; (iii) The nonlinearity requires that the fluctuations be relatively polarized with one another in the plane perpendicular to the equilibrium field. Thus, to fully capture the physics of the turbulent cascade one requires all three dimensions \citep{Tronko:2013,Howes:2015}. This requirement to retain all three dimensions to capture \Alfvenic\ turbulence persists in both full MHD and kinetic plasmas \citep{Schekochihin:2009,Howes:2015}. 

Finally, we highlight one other important fact about turbulence, which can be seen by examining the Elsasser energy equation, obtained by taking the dot product of $\Vz^\pm$ with \eqr{eq:NRElsasser} and integrating over all of space. Assuming periodic boundary conditions or that the fields vanish at infinity, one obtains
\begin{equation}\label{eq:NRenergy}
    \int d^3 \V{r} \frac{d |\Vz^\pm|^2}{dt} = 0.
\end{equation}
\eqr{eq:NRenergy} implies that there is no exchange of energy between the upward and downward propagating fluctuations, even during nonlinear interactions \citep{Maron:2001,Schekochihin:2009}. The collisions of counter-propagating fluctuations are therefore elastic: one wave packet can scatter another, but the individual energies of the $z^+$ and $z^-$ fluctuations do not change. 

\subsection{The Connection to Reduced MHD}\label{sec:RMHD}

As noted in \secref{sec:intro}, \Alfvenic\ turbulence proceeds in an anistropic fashion as it cascades to smaller scales, eventually leading to a state in which $k_\parallel \ll k_\perp$ (and $\delta B / B_0 \ll 1$) regardless of the initial isotropy of the plasma at the outer-scale. This scale-by-scale anisotropic turbulence cascade has been well observed in simulations \citep{Cho:2000,Maron:2001,Chen:2012} and in situ solar wind observations \citep{Horbury:2008,Wicks:2010a,Chen:2012,Chen:2016}. Thus, it would be advantageous to consider an ordering framework that leverages this fact. Fortunately, the minimal ordering assumptions for reduced MHD (RMHD) are anisotropic fluctuations ($k_\parallel \sim \epsilon k_\perp$), small amplitude fluctuations relative to the background (e.g., $\delta B \sim \epsilon B_0$), and characteristic timescales $\omega \sim k_\parallel v_A$, where $\epsilon \ll 1$. Note that early derivations of RMHD \citep{Kadomstev:1974,Strauss:1976} further assumed strong magnetization, implying plasma $\beta \ll 1$. However, more recent derivations \citep{Schekochihin:2007,Schekochihin:2009} have demonstrated that the RMHD equations are valid for arbitrary plasma $\beta${, assuming a homogeneous background}. Conveniently, the equations of RMHD are essentially identical to \eqr{eq:NRElsasser}, with the exception that the gradients in the nonlinear terms reduce to gradients perpendicular to the equilibrium magnetic field.

Much like the incompressible MHD equations, the fast wave is ordered out of RMHD. In the solar wind, this is well justified and supported by observations, which show that fast modes generally compose a small fraction of the solar wind \citep{Tu:1994,Howes:2012c,Klein:2012}. The RMHD equations have gained prominence as the preferred set of equations for describing \Alfvenic\ turbulence because they have a few favorable properties compared to incompressible MHD, despite their close similarity. First, the RMHD equations are a rigorous set of equations for describing collisional or collisionless \Alfvenic\ physics at scales large compared to the ion gyroradius, $k_\perp \rho_i \ll 1$ \citep{Schekochihin:2009}. Second, in the anisotropic limit of RMHD, the plus and minus \Alfven\ mode, plus and minus slow (or pseudo-\Alfven) mode, and the lone entropy mode cascades decouple. In other words, there are five independent cascade channels in RMHD, and the energy in each channel is independently conserved. Thus, the five channels do not exchange energy with one another, analogous to the two independent channels described above for incompressible MHD. The \Alfvenic\ cascade is described by the perpendicular vector components of \eqr{eq:NRElsasser}, the slow mode cascade by the parallel component, and the entropy mode cascade by the pressure balance condition \citep{Schekochihin:2009}. Formally, the RMHD equations are only valid for the slow and entropy modes in the collisional limit, because these modes are subject to collisionless wave-particle interactions via Landau \citep{Landau:1946} or Barnes \citep{Barnes:1966} damping, even for scales $k_\perp \rho_i \ll 1$. To describe these modes in the collisionless limit, one can instead use kinetic RMHD, which evolves the perpendicular dynamics using conventional RMHD and the parallel dynamics using the reduced ion kinetic equation \citep{Schekochihin:2009}. Third, in the anisotropic limit of RMHD, the \Alfven\ waves are not affected by the slow or entropy modes, and the slow and entropy modes do not cascade on their own. The slow and entropy modes behave as passive scalars, which can only be cascaded to small scales by interacting with \Alfven\ waves. This fact also implies that slow and entropy fluctuations will not be generated in situ in a purely \Alfvenic\ RMHD turbulence cascade \footnote{The passive scalar property of slow modes is only true in the RMHD limit. In fact, \cite{Tronko:2013} use incompressible MHD to demonstrate that in reduced dimensions, specifically in a two dimensional plane containing $\VB_0$, the slow mode cascade is the dominant cascade. Indeed, in this particular geometry, the \Alfvenic\ nonlinearity vanishes, leaving only a slow mode cascade.}.

Finally, before moving forward to discuss compressible MHD turbulence, we note that RMHD is equally valid for describing weak and strong (critically balanced) turbulence \citep{Galtier:2006b}. Comparing the strength of the nonlinear to linear terms in RMHD, we find
\begin{equation}
    \chi = \frac{\Vz^\mp\cdot\grad\Vz^\pm}{\V{v}_A \cdot\grad \Vz^\pm} \sim \frac{z^\mp k_\perp}{v_A k_\parallel} \sim \frac{k_\perp \delta B }{k_\parallel B_0}.
\end{equation}
The final expression is the ratio of asymptotically ordered quantities in RMHD and is therefore unordered. In other words, RMHD can equally well describe weak ($\chi \ll 1$) and critically balanced ($\chi \sim 1$) turbulence. 

As a brief aside, it is worthwhile at this point to provide a physical interpretation of critical balance, and justify why we have chosen to ignore the case $\chi > 1$. Physically, critical balance amounts to equating the linear, propagation time, $\tau_A = l_\parallel / v_A$, with the nonlinear (or ``eddy turnover'') time, $\tau_{NL} \sim l_\perp / u_\perp$. The case $\tau_A \ll \tau_{NL}$ ($\chi \ll 1$) is the weak turbulence case, which will eventually, at sufficiently small scales, transition to $\tau_A \sim \tau_{NL}$ because the weak turbulence cascade does not produce a cascade in the direction parallel to the equilibrium field: $l_\parallel$ is fixed at the outer-scale. Thus, in weak turbulence $\tau_A$ remains fixed while $\tau_{NL}$ will decrease with scale. The case $\tau_A \gg \tau_{NL}$  corresponds essentially to creating two dimensional structures perpendicular $\VB_0$ and is not sustainable at small scales separated from any external forcing. A fluctuation can only  remain two dimensional if it is causally connected, but for a system with an equilibrium $B_0$, the maximum parallel length over which a fluctuation can be coherent is $l_\parallel \sim \tau_{NL} v_A$, i.e., $l_\parallel / v_A \sim \tau_A \sim \tau_{NL}$. Thus, if a system is driven such that $\tau_A \gg \tau_{NL}$, it will rapidly relax back to critical balance, $\tau_A \sim \tau_{NL}$ at small scales. It is also worth noting that the critical balance condition, or predictions that follow from critical balance, have been observed in simulations \citep{Cho:2000,Maron:2001,Perez:2008,TenBarge:2012a,Mallet:2015,Mallet:2017b}, in situ solar wind observations \citep{Horbury:2008,Chen:2010a,Wicks:2010a,Chen:2012,Chen:2016}, and laboratory experiments \citep{Ghim:2013}  \footnote{Despite the arguments provided in favor of critical balance on physical grounds, as well as numerical and observational evidence, it remains a rather controversial topic in the field. For an opposing review, we refer the reader to \cite{Oughton:2020}.}.

Although the current discussion is focused on RMHD, which requires a strong equilibrium magnetic field, the preceding discussion concerning critical balance can equally well apply to small scales in a system without an equilibrium magnetic field. \Alfvenic\ fluctuations at a scale $l$ propagate along the total, local magnetic field at position $\V{x}_0$, $\VB_0^{local} (\V{x}_0) = \VB_0 + \sum_{l' \gtrsim l} \delta \VB_{l'}(\V{x}_0)$. Since the propagation and nonlinear times both decrease with scale, the fluctuations with $l' \gtrsim l$ are approximately static relative to small scale turbulence fluctuations, and those with $l' \lesssim l$ are more rapid and will average to zero. Therefore, small scale \Alfvenic\ turbulence always sees an approximately constant, local mean magnetic field.

\subsection{Compressible MHD}\label{sec:weakComp}

Given the complexity of incompressible turbulence, it is somewhat unsurprising that compressible MHD turbulence has received less attention. The first important point about compressible turbulence is the nature of the fast mode cascade. The fast mode propagates isotropically ($\omega_F \propto k$), and therefore the fast mode turbulence cascade is also isotropic, as confirmed in numerical simulations of compressible MHD \citep{Cho:2002a,Cho:2003}. In the weak limit, \cite{Chandran:2005,Luo:2006,Chandran:2008a} have examined the wave kinetic equation in detail to explore the couplings between the \Alfven\ and fast modes. For quasi-parallel fluctuations ($k_\parallel > k_\perp$), the frequency and wavevector matching conditions permit any of the following three-wave interactions on approximately equal footing: $\mathrm{A}+\mathrm{A}\rightarrow \mathrm{A}$, $\mathrm{A}+\mathrm{A}\rightarrow \mathrm{F}$, $\mathrm{A}+\mathrm{F}\rightarrow \mathrm{F}$, and $\mathrm{F}+\mathrm{F}\rightarrow \mathrm{F}$. However, in the obliquely propagating limit ($k_\parallel \lesssim k_\perp$), the cascades decouple, leaving only $\mathrm{A}+\mathrm{A}\rightarrow A$ and $\mathrm{F}+\mathrm{F}\rightarrow \mathrm{F}$, because in this limit, the frequency of the fast modes exceeds significantly that of the \Alfven\ modes, making the frequency matching conditions involving mixed wave modes impossible. Although the wavevector and frequency matching conditions are broadened as the turbulence becomes stronger, one still expects the turbulence to be dominated by interactions that are nearby in scale \citep{Kolmogorov:1941,Frisch:1995,Howes:2011c}, and thus also nearby in wavevector and frequency space. Considering these facts combined with the isotropic cascade of fast modes and the anisotropic cascade of \Alfvenic\ modes, it is expected that the cascades decouple at small scales, regardless of the initial wavevector distribution. In other words, regardless of the way a system is driven or initialized, the \Alfvenic\ portion of the cascade will eventually decouple and behave the same as the RMHD cascade described in the preceding section. Further, moderate amplitude fast modes rapidly form shocks as they propagate, and in weakly collisional plasmas, fast modes are moderately to strongly damped via resonant wave-particle interactions \citep{Landau:1946,Barnes:1966} for a wide range of plasma parameters \citep{Klein:2012}. Therefore, in the Newtonian limit, focusing only on the \Alfvenic\ turbulence cascade is, generally, well justified.

\section{The Equations of Relativistic Weak Turbulence}\label{sec:equations}
\subsection{Relativistic Elsasser Equations}\label{sec:relEls}

To begin our exploration of weak turbulence in relativistic, magnetically dominated plasmas, we would like to start from an equation set that: (i) makes direct contact with the Newtonian Elsasser equations, \eqr{eq:NRElsasser}, and (ii) highlights the fundamental role of \Alfven\ wave collisions in establishing the turbulence cascade. To this end, we turn to \cite{Chandran:2018}, who present a set of Elsasser-like equations for general relativistic MHD, including an inhomogeneous background. Here, we outline the derivation and consider only the special relativistic form with a homogeneous background. From this point forward, {we will employ Lorentz-Heaviside units and all speeds are normalized to the speed of light, $c=1$}, and we will assume a fixed Minkowski metric with signature $\eta^{\mu \nu} = \text{diag}\left\{-1,1,1,1\right\}$. The ideal relativistic MHD equations are mass conservation,
\begin{equation}\label{eq:Rcont}
    \partial_\nu(\rho u^\nu) = 0,
\end{equation}
the stress energy equation,
\begin{equation}\label{eq:RStress}
    \partial_\nu T^{\mu \nu} = 0,
\end{equation}
and the induction equation,
\begin{equation}\label{eq:RInduction}
    \partial_\nu (b^\mu u^\nu - b^\nu u^\mu) = 0,
\end{equation}
 where $\rho$ is the rest-mass density, $u^\mu = (\gamma,\gamma \Vv)$ is the four-velocity, $\gamma = 1 / \sqrt{1 - v^2}$ is the Lorentz factor. Furthermore,
\begin{equation}
    T^{\mu \nu} = \mathcal{E} u^\mu u^\nu + \left(p + \frac{b^2}{2}\right) \eta^{\mu \nu} - b^\mu b^\nu
\end{equation}
is the stress-energy tensor, 
\begin{equation}
    b^\mu = \frac{1}{2}\varepsilon^{\mu \nu \kappa \lambda} u_\nu F_{\kappa \lambda} = ( \gamma (\Vv \cdot \VB), \frac{1}{\gamma} \left[B^i + \gamma^2  (\Vv \cdot \VB) v^i \right])
\end{equation}
is the magnetic field four-vector, $b^2 = b^\mu b_\mu$, $F_{\kappa \lambda}$ is the Faraday tensor, $\mathcal{E} = h + b^2$, $h = \rho (1 + \epsilon) + p$ is the enthalpy density,  $\varepsilon$ is the specific internal energy, and $p$ is the gas pressure. Repeated indices indicate summation, Greek indices span the 4-dimensional spacetime (0 to 3), while Latin indices (1 to 3) correspond to the spatial directions in a suitably chosen 3+1 foliation of spacetime. We adopt an ideal equation of state $p = \rho \epsilon (\Gamma -1)$, with an adiabatic index $\Gamma = 4/3$ appropriate for relativistic plasmas. For this equation of state and adiabatic index, $h = \rho + 4 p$.

An Elsasser-like formulation of the relativistic MHD equations can be achieved by simply mulipltying \eqr{eq:RInduction} by $\pm \sqrt{\mathcal{E}}$, adding the result to \eqr{eq:RStress}, and dividing the sum by $\mathcal{E}$, yielding
\begin{equation}\label{eq:RElsasser}
    \partial_\nu (z^\mu_\pm z^\nu_\mp + \Pi \eta^{\mu \nu}) + (\frac{3}{4} z^\mu_\pm z^\nu_\mp + \frac{1}{4} z^\mu_\mp z^\nu_\pm + \Pi \eta^{\mu \nu})  \frac{\partial_\nu  \mathcal{E}}{\mathcal{E}} = 0,
\end{equation}
where
\begin{equation}
z^\mu_\pm = u^\mu \pm  \frac{b^\mu}{\sqrt{\mathcal{E}}}
\end{equation}
are the relativistic Elsasser four-vector fields, and
\begin{equation}
\Pi = \frac{2p + b^2}{2\mathcal{E}} = \frac{1}{2} - \frac{2 p + \rho}{2\mathcal{E}}.
\end{equation}

Much like in non-relativistic MHD, \eqr{eq:RElsasser} represents the evolution of upward and downward propagating Elsasser fields; however, \eqr{eq:RElsasser} represents the fully compressible system and is thus not completely analogous to \eqr{eq:NRElsasser}. In the relativistic limit, one cannot simply go to the incompressible limit typically used to derive incompressible MHD, because the maximum speed is the speed of light, and thus compressible fluctuations cannot be ordered in such a way as to instantaneously carry information away from a source. We can, however, consider a limit similar to  reduced MHD to isolate the \Alfvenic\ fluctuations described by \eqr{eq:RElsasser}. Such a limit, in which fluctuations are highly elongated relative to a mean magnetic field, is reasonable to consider for many astrophysical plasmas which often have strong mean fields, e.g., black hole accretion disks and jets \citep{Narayan:2003}, coronae \citep{Chandran:2018,Yuan:2019b}, and magnetar magnetospheres \citep{Parfrey:2012}. Further, as we will show in \secref{sec:asymptotic}, relativistic \Alfvenic\ turbulence proceeds in the same anisotropic sense as Newtonian turbulence, eventually leading to highly anisotropic, small amplitude fluctuations at small scales, regardless of the outer-scale conditions. Additionally, many astrophysical systems are characterized by exceptionally large inertial ranges\footnote{The range of scales well separated from driving, dissipation, and kinetic effects.}, vastly exceeding the three to four order of magnitude wide inertial range observed in the solar wind near Earth \citep{Bruno:2005,Howes:2008,Kiyani:2015,Chen:2016}.

\subsection{Relativistic Reduced MHD}\label{sec:RRMHD}

To derive a set of Elsasser equations for relativistic reduced MHD (RRMHD), we begin by separating the fluid into mean (background) quantities and fluctuating quantities of the form,
\begin{equation}
    b^\mu = \langle b^\mu \rangle + \delta b^\mu.
\end{equation}
We also construct an average fluid rest frame in which $\langle u^i\rangle$ vanishes\footnote{Note that $\langle u^\mu \rangle$ is not formally a four-velocity, since $\langle u^\mu \rangle \langle u_\mu\rangle = -(1 + \langle \delta v^2 \rangle)$.}. We define $\lambda$ to be the correlation length transverse to $\langle B^i \rangle$, and $L$ to be the characteristic scale of the background, i.e., the length scale parallel to $\langle B^i \rangle$. As in RMHD, we assume $\lambda / L \sim \order{\epsilon}$, where $\epsilon \ll 1$. In addition to the anisotropy assumption, we also assume that the characteristic frequency is of order the \Alfven\ frequency, $\partial_t \sim \langle B^i\rangle \partial_i \sim k_\parallel v_A$, and that the fluctuations are ordered small:
\begin{equation}
    \frac{\sqrt{\delta u^2}}{v_A} \equiv \frac{\sqrt{\delta u^\mu \delta u_\mu}}{v_A} \sim \sqrt{\delta b^2} \sim \order{\epsilon},
\end{equation}
\begin{equation}    
    \frac{\delta \rho}{\langle\rho\rangle} \sim \frac{\delta p}{\langle p\rangle} \sim \order{\epsilon^2},
\end{equation}
where 
\begin{equation}
    v_A^i = \frac{\langle b^i \rangle}{\sqrt{\langle\mathcal{E}\rangle}},
\end{equation}
and $v_A = \sqrt{v_A^i v_{Ai}}$. We will further assume that all background quantities, e.g., $\langle B^i\rangle = \VB_0, \langle\rho\rangle \equiv \rho_0,\langle p\rangle \equiv p_0$, are constant with no background inhomogeneities. 

As in Newtonian RMHD, in relativistic RMHD, the fast wave will be ordered out of the system, since $\omega_F \sim k \gg \omega \sim k_\parallel$. RRMHD will describe \Alfven\ and pseudo-\Alfven\ fluctuations; however, we are not interested in the pseudo-\Alfven\ waves, which have fluctuations parallel to the background magnetic field and are a separate, passive, cascade channel. Therefore,  we will also assume $B_{0i} \delta z^i_\pm = \order{\epsilon^2}$ to remove the pseudo-\Alfven\ modes. With all of the above assumptions, we note that $\gamma \sim 1 + \order{\epsilon^2}$, $\partial_i \delta u^i \sim \partial_\mu \delta z^\mu_\pm \sim \langle b^t\rangle \sim \delta b^t \sim \delta u^t / v_A \sim \order{\epsilon^2}$.

Applying the above reduced assumptions to the relativistic Elsasser equation, \eqr{eq:RElsasser}, we arrive at
\begin{equation}\label{eq:RRElsasser}
    v^\nu_{A\mp} \partial_\nu \delta z^\mu_\pm  = -\delta z^\nu_\mp \partial_\nu \delta z^\mu_\pm - \partial_\nu (\delta \Pi \eta^{\mu \nu}),
\end{equation}
where to lowest non-zero order
\begin{equation}
    v_{A\pm}^\mu = \left(1,\pm\frac{\VB_0}{\sqrt{\mathcal{E}_0}}\right)  \equiv (1,\pm \Vv_A),
\end{equation}
\begin{equation}
\delta z^\mu_\pm = \delta u^\mu \pm  \frac{\delta b^\mu}{\sqrt{\mathcal{E}_0}} = \left(0,\delta \Vv_\perp \pm \frac{\delta \VB_\perp}{\sqrt{\mathcal{E}_0}}\right),
\end{equation}
\begin{equation}
    \delta \Pi = -\frac{2 \delta p + \delta \rho}{2 \mathcal{E}_0} + \frac{2 p_0 + \rho_0}{2 \mathcal{E}_0}\frac{\delta \mathcal{E}}{\mathcal{E}_0},
\end{equation}
 $\mathcal{E}_0 = 4 p_0 + \rho_0 + B_0^2$, and $\delta \mathcal{E} = 4\delta p + \delta \rho + 2 B_0 \delta B_\parallel$. In three-vector form, the equation set is particularly simple and similar to the Newtonian Elsasser equations,
\begin{equation}\label{eq:RRElsasser3}
    \pfrac{\delta \Vz_{\perp \pm}}{t} \mp \Vv_A \cdot \grad \delta \Vz_{\perp \pm} = - \delta \Vz_{\perp \mp} \cdot \grad_\perp \delta \Vz_{\perp \pm} - \grad_\perp \delta \Pi.
\end{equation}
The terms of the left-hand side of \eqr{eq:RRElsasser} are linear, while those on the right-hand side are nonlinear.  Due to the assumptions regarding pressure and parallel fluctuations, the only relevant components of \eqr{eq:RRElsasser} are the two components transverse to the mean magnetic field, as expressed in \eqr{eq:RRElsasser3}. The other two components, time-like and parallel, are one order higher in the ordering parameter, $\epsilon$. {Importantly, the parallel fluctuations do not appear at lowest order in the nonlinear $\V{E} \times \VB$ term in the perpendicular, \Alfven\ wave equations, \eqr{eq:RRElsasser} and \eqr{eq:RRElsasser3}, and they therefore do not cascade the \Alfven\ waves. However, the primary nonlinear term for the parallel fluctuations is of the form $\delta \Vz_{\perp_\pm} \cdot \grad_\perp \delta z_\parallel$. Thus, the parallel fluctuations are passively scattered/mixed by the \Alfven\ waves, much like in Newtonian RMHD turbulence \citep{Schekochihin:2009}.} Note that as in the Newtonian RMHD equations, the system is closed by taking the divergence of \eqr{eq:RRElsasser3}, or four-divergence of \eqr{eq:RRElsasser}, to find an equation for $\delta \Pi$.

 
 The reduced relativistic Elsasser equations, \eqr{eq:RRElsasser3}, for relativistic RMHD are  identical in form to the Newtonian RMHD equations, and at this point, the standard approach to solve for the \Alfven\ dynamics would be to take the curl of \eqr{eq:RRElsasser3} to eliminate the pressure term and solve for the Elsasser potentials rather than the Elsasser fields. Indeed, this is the approach we will also take; however, we note that it is possible to simplify the system even further, because there is a straightforward limit to consider in relativistic plasmas. This final simplification to consider is to remove the pressure fluctuations by assuming that we are in the magnetically dominated regime, $\sigma = b^2 / h \gg 1$. In this limit, $\langle \Pi \rangle = 1/2$, and $\delta \Pi \sim \order{\epsilon^2} / \sigma \ll \order{\epsilon^2}$. Therefore, we arrive at  the magnetically dominated, relativistic, reduced MHD equations
 \begin{equation}\label{eq:RMRElsasser}
    v^\nu_{A\mp} \partial_\nu \delta z^\mu_\pm  = -\delta z^\nu_\mp \partial_\nu \delta z^\mu_\pm.
\end{equation}
In three-vector form, the equation set is particularly simple,
\begin{equation}\label{eq:RMRElsasser3}
    \pfrac{\delta \Vz_{\perp \pm}}{t} \mp  \grad_\parallel \delta \Vz_{\perp \pm} = - \delta \Vz_{\perp \mp} \cdot \grad_\perp \delta \Vz_{\perp \pm}.
\end{equation}
Being in the $\sigma \gg 1$ limit, \eqr{eq:RMRElsasser} and \eqr{eq:RMRElsasser3}, are essentially the reduced analog of the force-free limit of the ideal MHD equations \citep{Gruzinov:1999,Komissarov:2002}. In this limit, $v_A = c = 1$.

To arrive at this simple form, it may seem that we have rather seriously brutalized the relativistic MHD equations by applying a series of restrictive asymptotic orderings:  (i) $k_\parallel \ll k_\perp$; (ii) $\omega \sim k_\parallel v_A$; (iii) Small amplitude fluctuations with constant (mean) backgrounds; (iv) Second order parallel and pressure fluctuations; (v) Magnetically dominated, $\sigma \gg 1$. The reduced assumptions, (i)-(iii), and (v) are consistent with one another, i.e., they can all be obtained by assuming the system is strongly magnetized. 
{In the Newtonian limit, the fast mode can be eliminated by assuming the system is incompressible; however,} the finite speed of light prevents easy elimination of the fast mode \citep{Takamoto:2017} without assuming the fluctuations are anisotropic ($k_\parallel \lesssim k_\perp$). To remove the slow mode, we assumed that parallel and pressure fluctuations are second order quantities when deriving the reduced relativistic Elsasser equations. Moving to the magnetically dominated regime produces the same result, since in the $\sigma \rightarrow \infty$ limit, the slow wave is also ordered out of the system. Thus, despite the myriad assumptions to achieve a simple set of equations for describing relativistic \Alfvenic\ turbulence, they are all self-consistent. This set of assumptions is also relevant to many astrophysical systems, such as magnetars \citep{Li:2015,Li:2019b,Yuan:2020}, glitches affecting radio emission from pulsar magnetospheres \citep{Bransgrove:2020,Yuan:2020b}, and X-ray emitting coronae around black hole accretion disks \citep{Thompson:1998,Chandran:2018}. We also note that the most important assumption is wavevector anisotropy, $k_\parallel \ll k_\perp$,  which naturally arises as the \Alfvenic\ turbulence cascades to small scales.

\subsection{Connection and Comparison to Newtonian Reduced MHD Solutions}
{
\eqr{eq:RRElsasser3} will form the basis for the following analysis in \secref{sec:asymptotic}, because it represents the simplest set of equations for describing \Alfvenic\ turbulence in magnetized, relativistic environments and has a form that is nearly identical to the Newtonian RMHD Elsasser equations. Thus, the system shares many properties with the Newtonian RMHD system of equations, some of which can be seen immediately: (i) The system supports linear \Alfven\ modes, which require $k_\parallel \ne 0$ to propagate; (ii) The nonlinearity requires counter-propagating fluctuations; (iii) The nonlinearity requires that the fluctuations be polarized with respect to each other in the perpendicular plane. Further, the wave kinetic equations for the system coincide with the kinetic equations for shear \Alfven\ waves derived in \cite{Galtier:2000} in the incompressible MHD limit and \cite{Boldyrev:2009a} in the RMHD limit neglecting imbalanced interactions, i.e., non-zero cross helicity. As demonstrated in these papers, the wave kinetic equation evolves the spectral energy $e^{\pm} = \langle |\delta \Vz_\perp^\pm (\mathbf{k})|^2 \rangle$ and can be expressed as
\begin{equation}
    \frac{\partial e^\pm(\mathbf{k})}{\partial t} = \int M_{k,pq} e^\mp(\mathbf{q}) [e^\pm(\mathbf{p}) - e^\pm(\mathbf{k})] \delta(q_\parallel)d_{k,pq},
\end{equation}
where  $M_{k,pq} = (\pi / v_A) (\mathbf{k}_\perp \times \mathbf{q}_\perp)^2 (\mathbf{k}_\perp \cdot \mathbf{p}_\perp)^2 / (k_\perp q_\perp p_\perp)^2$ and $d_{k,pq} = \delta(\mathbf{k} - \mathbf{p} - \mathbf{q})d^3p \ d^3q$. Since the kinetic equation is unchanged from RMHD, we also know that the weak turbulence solutions: (i) are dominated by three-wave interactions; and (ii) lead to an energy spectrum of the form $f(k_\parallel) k_\perp^{-2}$, where $f(k_\parallel)$ is set by external forcing or initial conditions. In the following sections, we explore in detail the three-wave interaction of \Alfven\ waves first analytically via heuristic solutions through third order, and then numerically to demonstrate the decoupling of the fast mode from the \Alfven\ waves in the obliquely propagating (reduced) limit.  Note that although we employ the RRMHD equations, \eqr{eq:RRElsasser3}, to derive the turbulence solutions in the following section, the \Alfven\ solutions are identical for the $\sigma \rightarrow \infty$ limit of the RRMHD equations, \eqr{eq:RMRElsasser3}.
}

\section{Three-wave Weak \Alfvenic\ Interactions}\label{sec:asymptotic}
To construct {analytical} solutions, we now consider a subsidiary  expansion of the form $\zeta_\pm = \varepsilon \zeta_{1\pm} +  \varepsilon^2  \zeta_{2\pm} + \cdots$, where $\delta \Vz_\pm = \hz \times \grad_\perp \zeta_\pm$ defines the Elasser potential, $\zeta$, and $\zeta_{i \pm}(t=0) = 0$ for $i > 1$. Note that since the relativistic equations are identical in form to the Newtonian limit, the solution to the RRMHD equations is also the same, and the full solutions appear in \secref{sec:appA}. Therefore, we primarily summarize the solutions found in \cite{Howes:2013a}; however, we note that we include corrections to \cite{Howes:2013a} Eqs. (22), (28-29), and all equations involving $\zeta_{3-}$, including the equations for $\VB_{\perp3}$ and $\V{E}_{\perp3}$. These errors have been corrected in the \secref{sec:appA}, and the changes are highlighted in red. 
 
For specificity, we will consider a periodic domain of size $L_x \times L_y \times L_z$ with $\VB_0 = B_0 \hz$. At $t=0$, we will initiate two counter-propagating, perpendicularly polarized fluctuations
\begin{equation}\label{eq:ICs}
    \begin{split}
        \Vz_{1+} &= z_+ \cos{(k_\perp x - k_\parallel z - \omega_0 t)}\hy\\
        \Vz_{1-} &= z_- \cos{(k_\perp y + k_\parallel z - \omega_0 t)}\hx,
    \end{split}
\end{equation}
where $z_\pm$ are the initial amplitudes, $k_\perp = 2\pi / L_x = 2 \pi / L_y$, and $k_\parallel = 2\pi / L_z$. We maintain the convention that $k_\perp$, $k_\parallel$, and $\omega_0$ are positive constants, and the direction of propagation is supplied by the explicit sign of $k_\parallel$. To describe the wave modes that arise from the nonlinear evolution, we will use the notation $(k_x/k_\perp, k_y/k_\perp, k_z / k_\parallel)$. For instance, the plus and minus initial wave modes in this notation are $(1,0,-1)$ and $(0,1,1)$ respectively.

The solutions through third order are lengthy, and as such can be found in \secref{sec:appA}. Here, we summarize the important findings at each order. The initial conditions provided in \eqr{eq:ICs} satisfy the lowest order equations if $\omega_0$ is the linear \Alfven\ wave frequency, $\omega_0 = k_\parallel v_A$. Thus, at lowest order the solution described counter-propagating, linear \Alfven\ waves.

The second order solutions for the electric and magnetic field are given by Eqs. \eqref{eq:O2B} and \eqref{eq:O2E}, from which we can immediately see that at second order, all components are purely oscillatory, i.e., there is no secularly growing mode. We can also see that two Fourier modes are generated at this order, $(1,1,0)$ and $(-1,1,2)$. Both of these modes satisfy the wavevector matching conditions, which require $\V{k}_2 = \V{k}_{1-} \pm \V{k}_{1+} = (0,1,1) \pm (1,0,-1)$. The $(-1,1,2)$ mode satisfies the conditions for a linear \Alfven\ wave. Specifically, the frequency of this mode is $\omega = \pm2\omega_0 = \pm 2 k_\parallel v_A = \pm k_z v_A$, and the mode obeys the \Alfven\ wave eigenrelation
\begin{equation}\label{eq:AWEigen}
    \frac{\delta \VB_\perp}{\sqrt{\mathcal{E}_0}} = \pm \delta \Vu_\perp = \pm \frac{\V{E}_\perp}{B_0} \times \hz.
\end{equation}
These two linear modes are counter-propagating along the background magnetic field; however, they are not perpendicularly polarized. Therefore, their interaction is simply a linear superposition that forms a standing wave for this particular symmetric initial condition. The $(1,1,0)$ mode is a purely magnetic mode that has no structure along the background field ($k_z = 0$) but oscillates with frequency $\omega = 2 \omega_0$. This mode is not a linear \Alfven\ mode and corresponds to a nonlinear magnetic shear. The interaction of this shear mode with the initial wave modes is the nonlinear interaction that will provide secular growth at the next order.

As with the second order equations for $\VB_\perp$ and $\V{E}_\perp$, we can once again straightforwardly interpret the third order solutions, Eqs. \eqref{eq:O3B} and \eqref{eq:O3E}. First, we note that there are now secularly (proportional to $t$) growing modes which are boxed for clarity: $(2,1,-1)$ and $(1,2,1)$. Both of these modes are linear \Alfven\ waves: They have frequency $\omega = \pm \omega_0 = \pm k_\parallel z = \pm k_z z$ and obey the \Alfven\ wave eigenrelations, \eqr{eq:AWEigen}. Therefore, the modes correspond to linear \Alfven\ waves propagating up, $(1,2,1)$, and down, $(2,1,-1)$, the background magnetic field. These secularly growing modes are phase-shifted relative to the initial modes by $-\pi / 2$, have perpendicular wavevectors $\sqrt{5} k_\perp$, but $|k_z| = k_\parallel$. Therefore, the weak turbulence cascade proceeds to smaller scales in the perpendicular plane, but the parallel scales are conserved, confirming the prediction from simple three-wave matching conditions. Further, as alluded to in the previous section, these secular modes follow from the interaction of the second-order $(1,1,0)$ mode with the  initial \Alfven\ waves: $\V{k}_3 = \V{k}_{2} + \V{k}_{1\pm} = (1,1,0) + (1,0,-1) = (2,1,-1)$ and $(1,1,0) + (0,1,1) = (1,2,1)$, i.e., each third-order mode conserves not just the magnitude but also the sign of $k_z$. This fact confirms that there is no exchange of energy between upward and downward propagating fluctuations. 

The other six purely oscillatory components are a mixture of linear \Alfven\ waves and non-linear structures. Unlike the second-order solutions, there is not a purely magnetic mode at third-order. Four of the components, those with wavevectors $(2,1,-1), (1,2,1), (-2,1,3)$, and $(-1,2,3),$ have $\sqrt{5} k_\perp$, but these all have both linear and non-linear components. Similarly, the $(0,1,1)$ and $(1,0,-1)$ components appear as both linear and non-linear terms. Note that these final two components have the same wavevector as the initial \Alfven\ waves, but these third-order modes have different phases and will serve to cancel the initial modes.


\section{Numerical Comparison}\label{sec:simulations}
\subsection{Simulation Description}\label{sec:sim_setup}

To confirm the analytical results of \secref{sec:equations} and \secref{sec:asymptotic}, we consider the nonlinear interaction between two perpendicularly polarized \Alfven\ waves that counter-propagate in a 3D, periodic domain along a uniform guide field $\mathbf{B_0} = B_0 \mathbf{\hat{z}}$ using the general relativistic MHD code \BHAC\ \citep{Porth:2017,Olivares:2019,Ripperda:2019,Ripperda:2019a}. The recently added force-free limit for the resistive MHD code \BHAC\ employs the numerical scheme of \cite{Ripperda:2019,Ripperda:2019a} and damps force-free violations $E>B$ and $\V{E} \cdot \V{B} \neq 0$ on resistive time scales\footnote{Full details of the simulation code, including convergence, numerical diffusion, and numerical dispersion studies, can be found in Paper II \citep{Ripperda:2021}.}. We employ both a period cubic domain with $L_{\perp}=L_x=L_y=L_{\parallel} = L_z = 2\pi$ and a periodic elongated domain with $L_{\parallel} = 10 L_\perp = 20\pi$, with resolution $N_x = N_y = N_z = 256$ and $N_z = 2560$ cells for the elongated case. Initially, we set a gas-to-magnetic-pressure ratio of $\beta=2p/B_0^2=0.02$ and magnetization $\sigma_{cold} \equiv b^2 / \rho \in [0.01;0.1;1;10;100]$ (corresponding to $\sigma \in [0.01;0.1;1;7;20]$), where we vary the density, $\rho$, but maintain a constant guide field, $\mathbf{B_0} = \mathbf{\hat{z}}$, and pressure $p=0.01$ with adiabatic index $\Gamma=4/3$ for an ideal relativistic gas. In the force-free case, $\beta \rightarrow 0$, $\sigma \rightarrow \infty$, and $v_A \rightarrow 1$.

As in \secref{sec:asymptotic}, we prescribe a scale-free definition of characteristic wavelengths $(k_x/k_\perp, k_y/k_\perp,$ $k_z / k_\parallel)$, and initialize our \Alfven\ wave simulations with two overlapping, counter-propagating, and perpendicularly polarized \Alfven\ waves described by the wavevectors $\mathbf{k}_{+}=(1,0,-1)$, and $\mathbf{k}_{-}= (0,1,1)$. 
The magnetic field is initialized through a vector potential $\mathbf{A} = (-B_0 y, 0, \delta B_\perp [\sin(k_\perp x+ k_\parallel z) + \sin(k_\perp y-k_\parallel z)])$, representing the initial counter-propagating \Alfven\ waves with frequency $\omega_0 = k_{\parallel} v_A $. The electric field is initialized as $\mathbf{E} = (v_A B_y, v_A B_x,0)$ such that the velocity is equal to the drift velocity, $\delta \Vu_\perp = \mathbf{E} \times \mathbf{B}/B^2$. Note that the overlapping \Alfven\ waves, in contrast to a single \Alfven\ wave, are not an exact force-free equilibrium due to a small second-order violation $\mathbf{E}_{\mp} \cdot \mathbf{B}_{\pm} \neq 0$ between the fields of the two waves, which is damped on a short time-scale in \BHAC.

The strength of the nonlinearity is characterized by $\chi = k_{\perp} \delta B_{\perp} / k_{\parallel} B_0$. To maintain weak turbulence, we fix $\chi = 0.01$ in all of the following simulations. By fixing $\chi$, we expect to maintain self-similar behavior as we explore elongated domains that approximate the reduced limit. Thus, in the cubic domain with $k_{\perp} = k_{\parallel}$, $\delta u_{\perp} / v_A = \delta B_\perp / B_0 = 0.01$, and in the elongated case with $k_{\perp} = 10 k_{\parallel}$, $\delta u_{\perp} / v_A = \delta B_\perp / B_0 = 0.001$.

\subsection{\Alfven\ Wave-\Alfven\ Wave Collisions}\label{sec:aw-aw}

\subsubsection{Cubic Domain}\label{sec:awawcubic}

\begin{figure}
  \centering
  \includegraphics[width=0.49\textwidth]{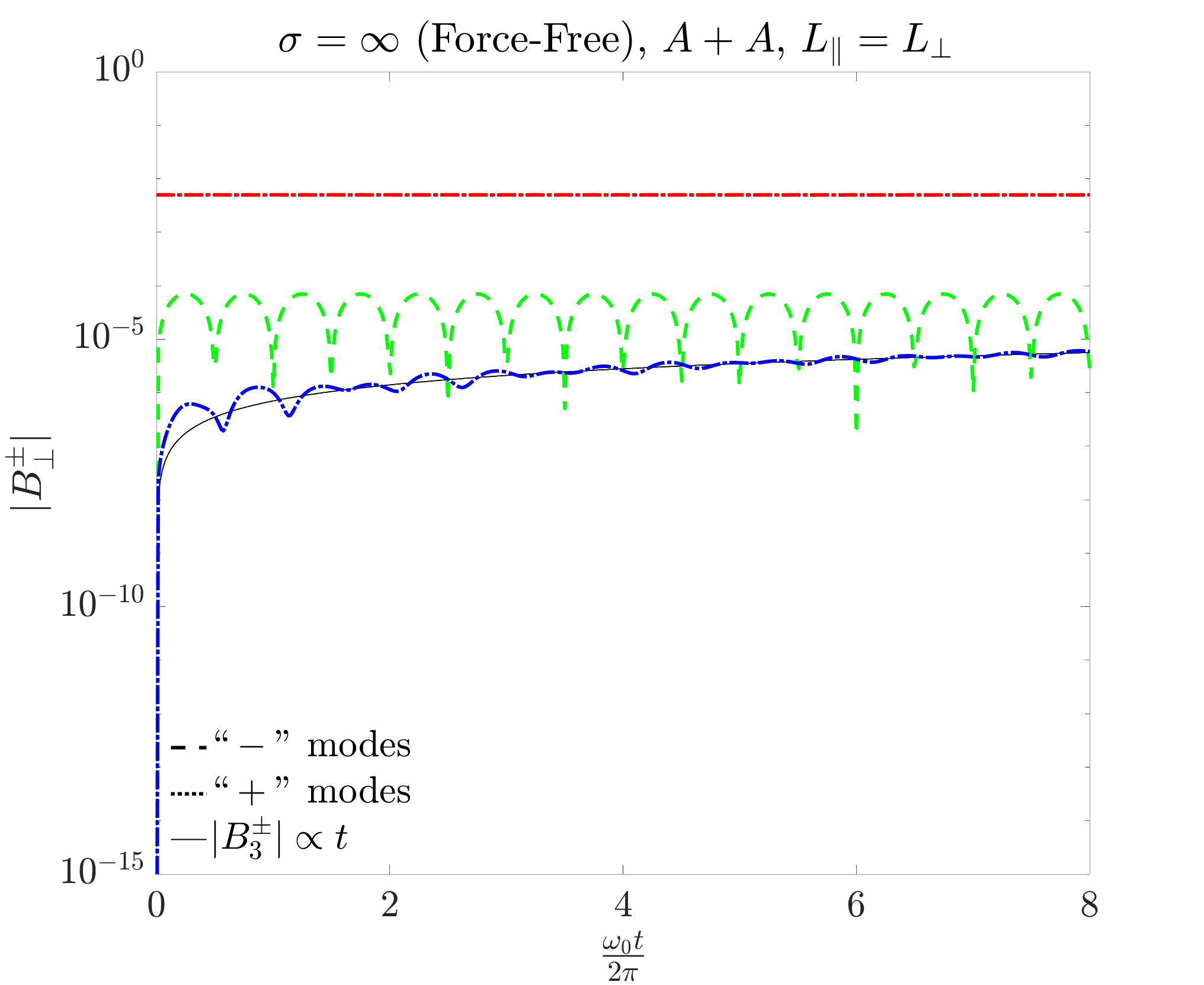}
  \includegraphics[width=0.49\textwidth]{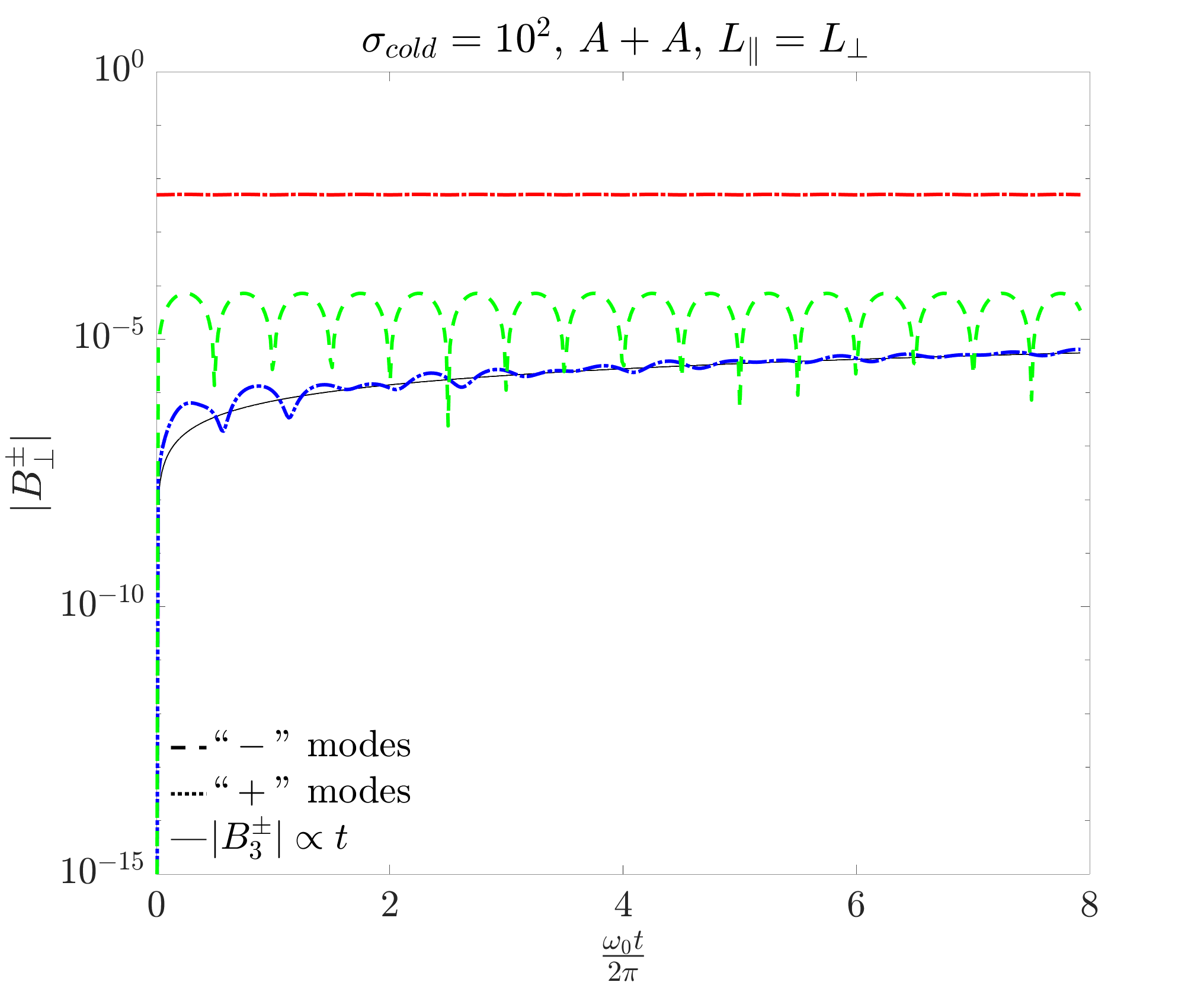}
  
  \includegraphics[width=0.49\textwidth]{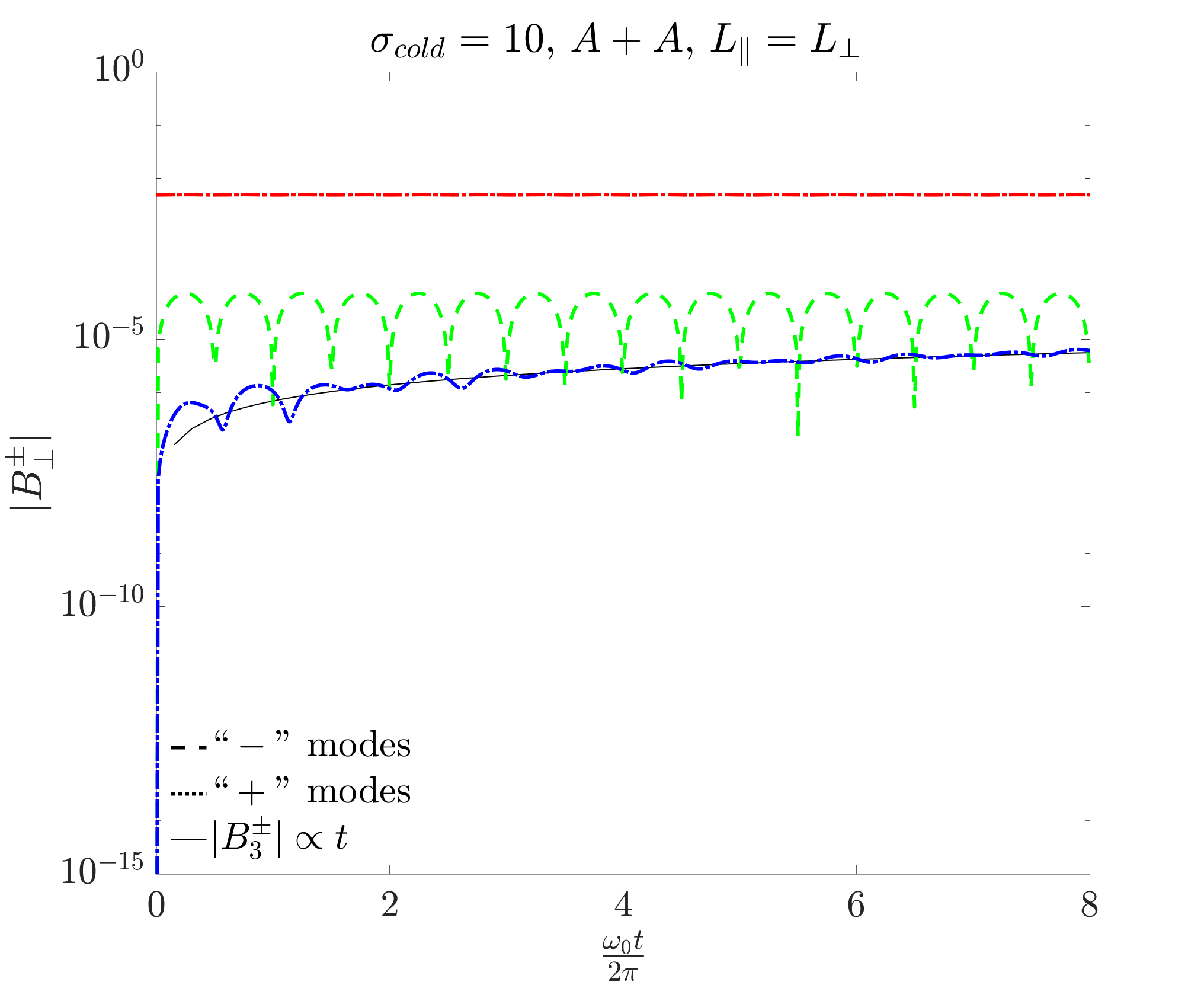}
  \includegraphics[width=0.49\textwidth]{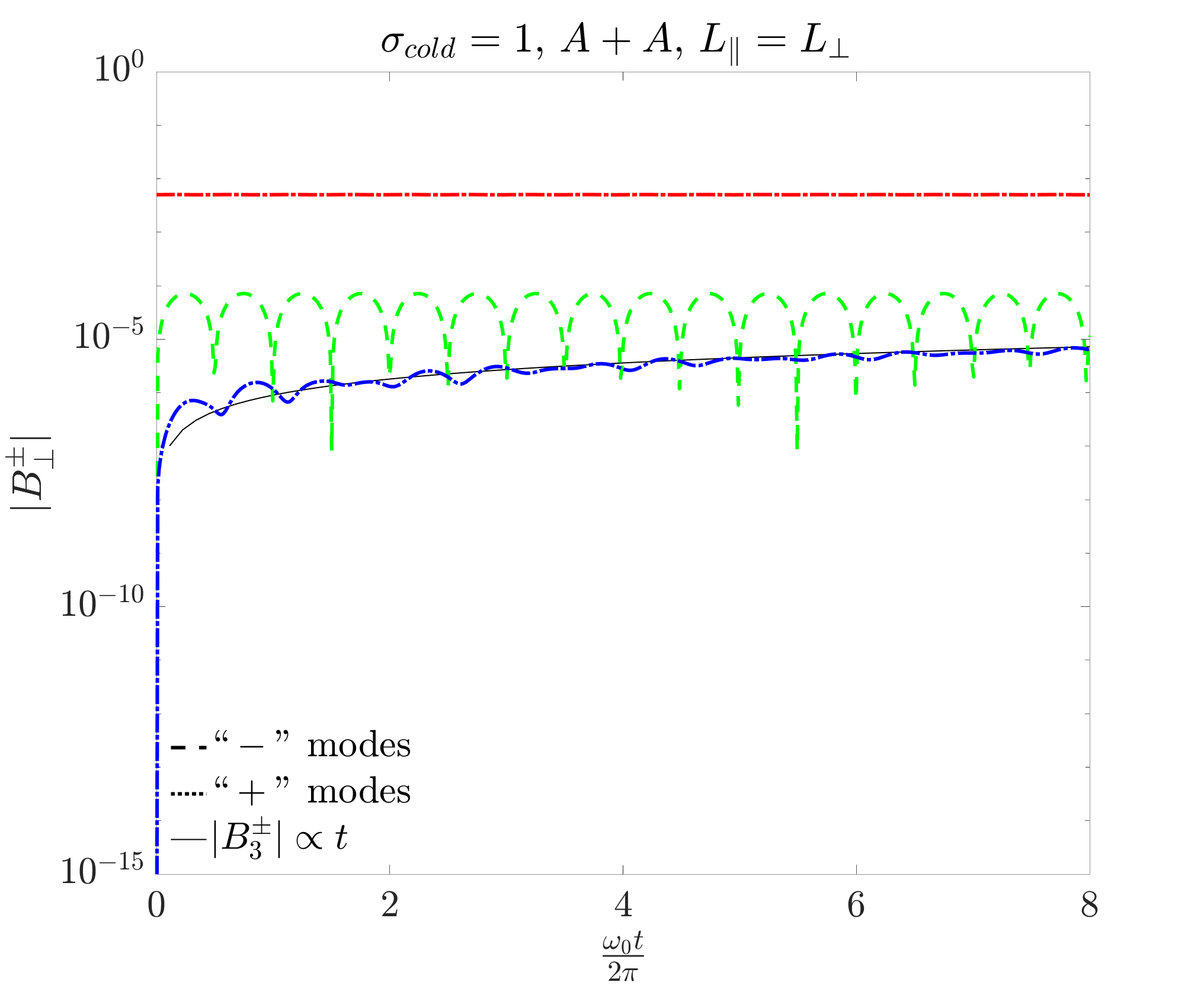}
  
  \includegraphics[width=0.49\textwidth]{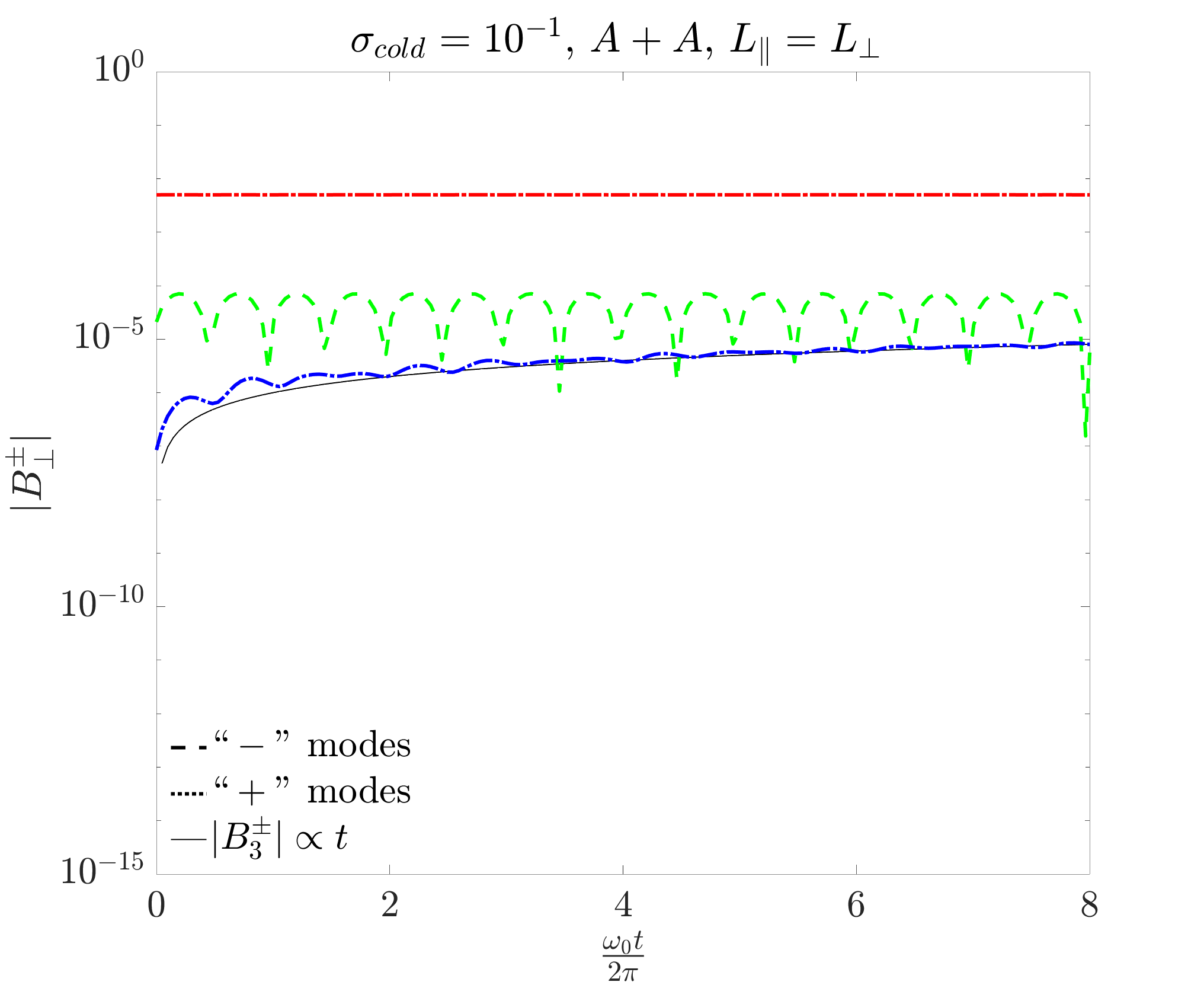}
  \includegraphics[width=0.49\textwidth]{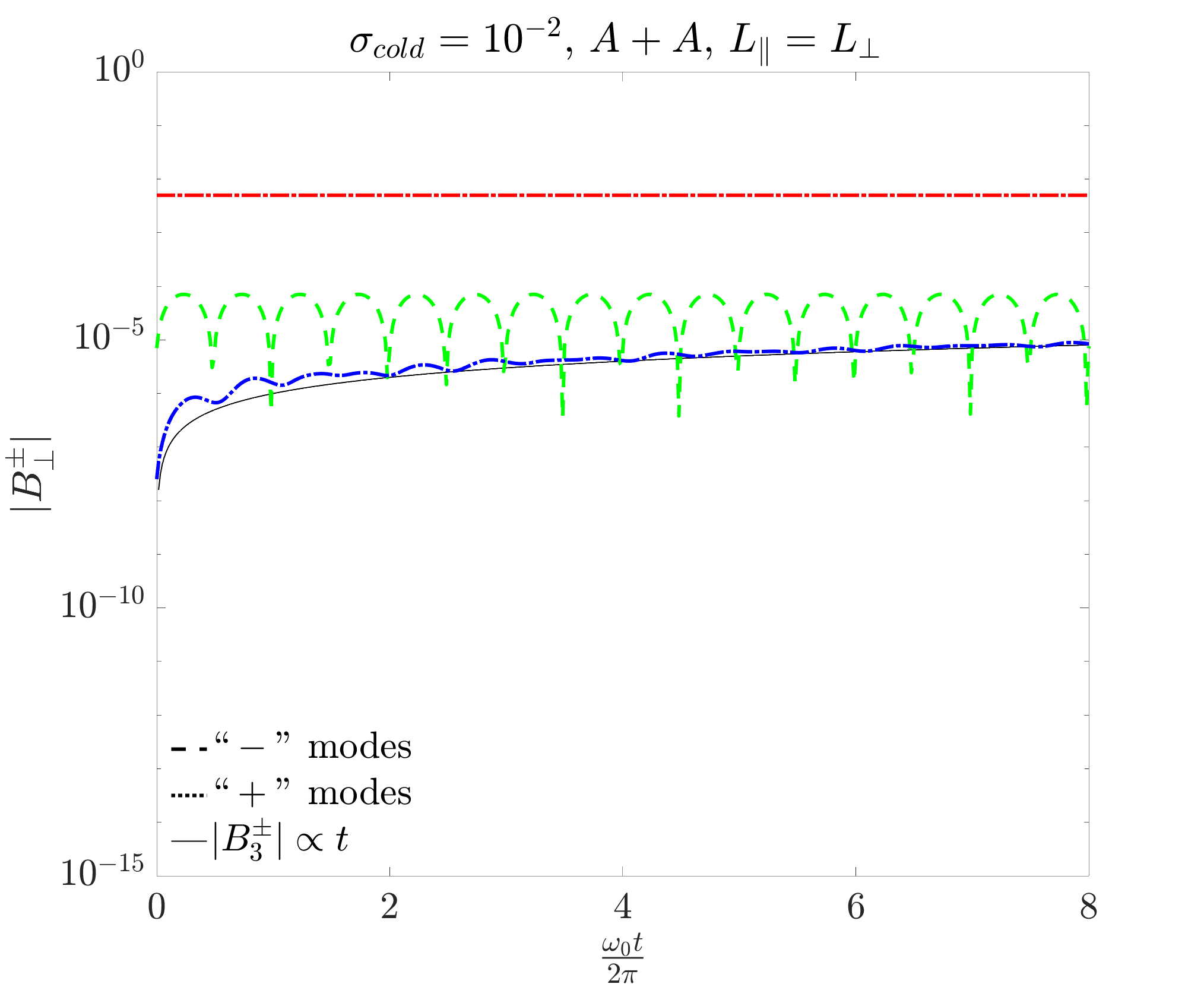}
  \caption{Mode evolution of $B_{\perp}$ for a cubic domain  with $\chi=0.01$ and $\sigma_{cold} \equiv b^2 / \rho \in [0.01;0.1;1;10;100;\infty]$, presented in descending order of $\sigma$ from top-left to bottom-right. The red lines correspond to the initial counter-propagating \Alfven\ modes: the $(0,1,-1)$ and $(1,0,1)$ modes are shown by dashed and dotted lines, respectively. Dashed green lines correspond to the second order nonlinear shear mode, $(1,1,0)$. The $(1,2,-1)$ and $(2,1,1)$  third order, secularly growing modes are shown by dashed and dotted blue lines, respectively. The black line corresponds to a secular growth directly proportional to $t$ {and scaled by the final amplitude of the third order mode}.}
\label{fig:modesBperp}
\end{figure}

In \figref{fig:modesBperp}, we present the evolution of the mode amplitudes of the $B_\perp$ (\Alfvenic) fluctuations in a cubic domain scanning $\sigma_{cold} \equiv b^2 / \rho \in [0.01;0.1;1;10;100;\infty]$, presented in descending order of $\sigma$ from top-left to bottom-right. The initial, counter-propagating, and perpendicularly polarized \Alfven\ waves are represented by the red lines in each figure, the $(0,1,-1)$ mode by dashed lines and the $(1,0,1)$ mode by dotted lines. These modes interact to produce a secondary, nonlinear, magnetic shear mode indicated by the green line, representing the $(1,1,0)$ mode. Note that this mode is purely oscillatory in time with $\omega = 2 \omega_0$, as found in \secref{sec:order2}. Finally, the secondary mode interacts with each of the primary \Alfven\ waves to produce at third order two higher $k_\perp$ \Alfven\ waves, but with the same $k_\parallel$ as the primary waves, where the $(1,2,-1)$ and $(2,1,1)$ modes are represented by the blue dashed and dotted curves. These dynamic are in agreement with the results of \secref{sec:order3}. These modes grow secularly in time, $B_{\perp3} \propto t$, as indicated by the black line in each panel{, which is a curve proportional to time and scaled by the final amplitude of $B_{\perp3}$}. The net result of this \Alfven\ wave-\Alfven\ wave collision is the anisotropic transfer of energy from large to small scales, which is governed by three-wave interactions.

\subsubsection{What About Fast Waves? Exploring the \Alfven\ Wave-Fast Wave Coupling}\label{sec:fastaw}

Our analytical analysis in the preceding sections purposefully neglected the fast wave by choosing the reduced limit. However, the fast wave may play an important role in releasing energy in strongly magnetized, relativistic, astrophysical plasmas, because the fast wave can travel across field lines \citep{Li:2019b,Yuan:2020}. For instance, \Alfven\ waves travel along magnetic fields, and on closed magnetic field lines in pulsar and magnetar magnetospheres, their energy is confined to the magnetosphere, but fast waves can release their energy into the surrounding medium by propagating across field lines. For this reason, the coupling between the fast and \Alfven\ branches in relativistic MHD and its force-free limit has been explored in detail in recent years \citep{Cho:2005,Takamoto:2016,Takamoto:2017,Li:2019b}. Through a numerical simulation study of relativistic turbulence, \cite{Takamoto:2016,Takamoto:2017} find that the fast-to-\Alfven\ mode power scales as $(\delta u_f / \delta u_A)^2 \propto \sqrt{(1+\sigma)} \delta u_A / c_{f\perp}$ in the $\sigma \gtrsim 1$ limit, where $c_{f\perp}$ is the fast mode speed in the perpendicular plane. In the $\sigma \ll 1$ limit, \cite{Cho:2002a,Cho:2003} find that $(\delta u_F / \delta u_A)^2 \propto \delta u_A / c_{f\perp}$. Thus, for $\sigma \ll 1$, the fast and \Alfven\ branches decouple as the background magnetic field strength is increased; however, for $\sigma \gtrsim 1$, the coupling between the modes increases with $\sqrt{1 + \sigma}$, asymptotically approaching  unity for large $\sigma$. 

\begin{figure}
  \centering
  \includegraphics[width=0.49\textwidth]{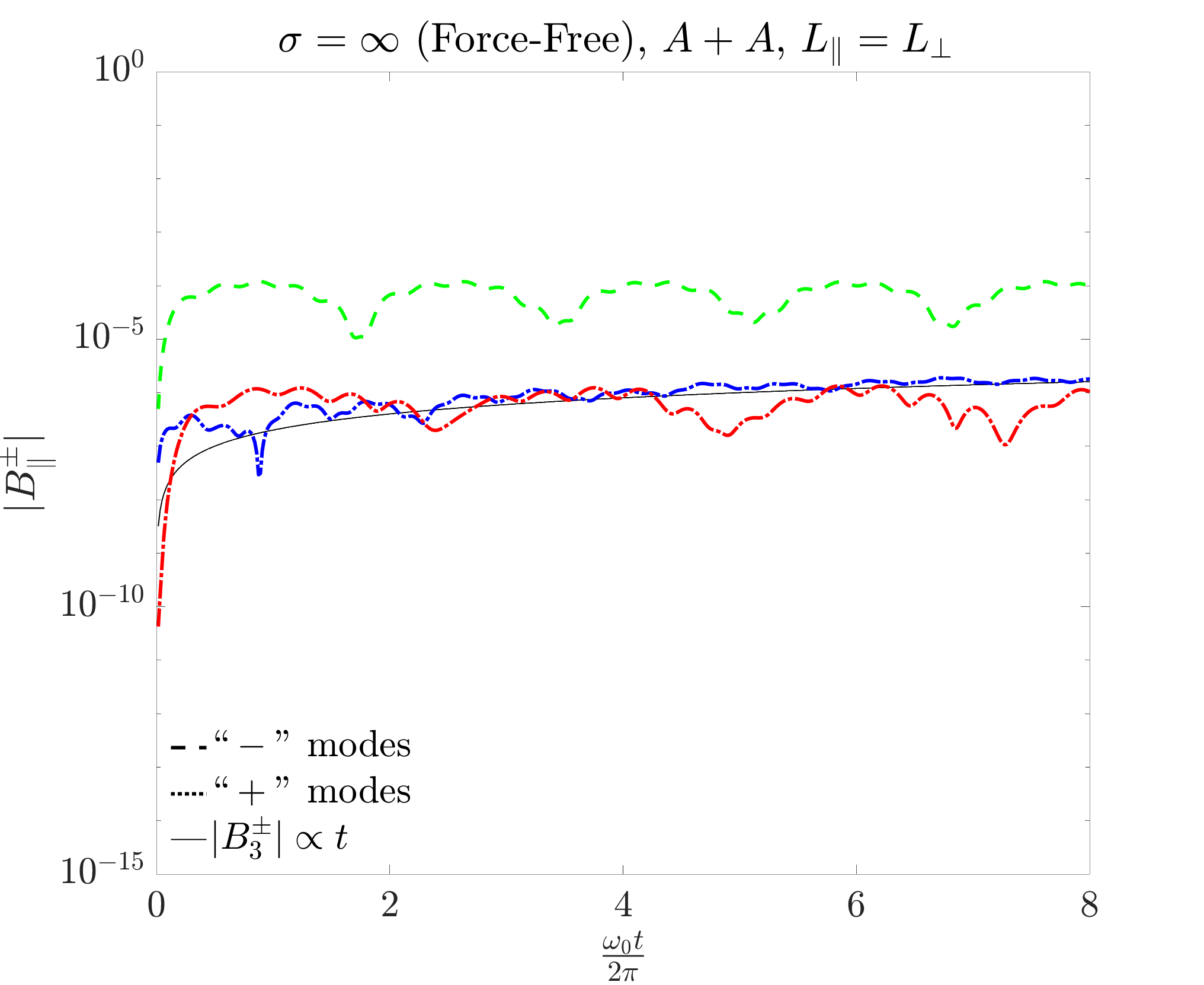}
  \includegraphics[width=0.49\textwidth]{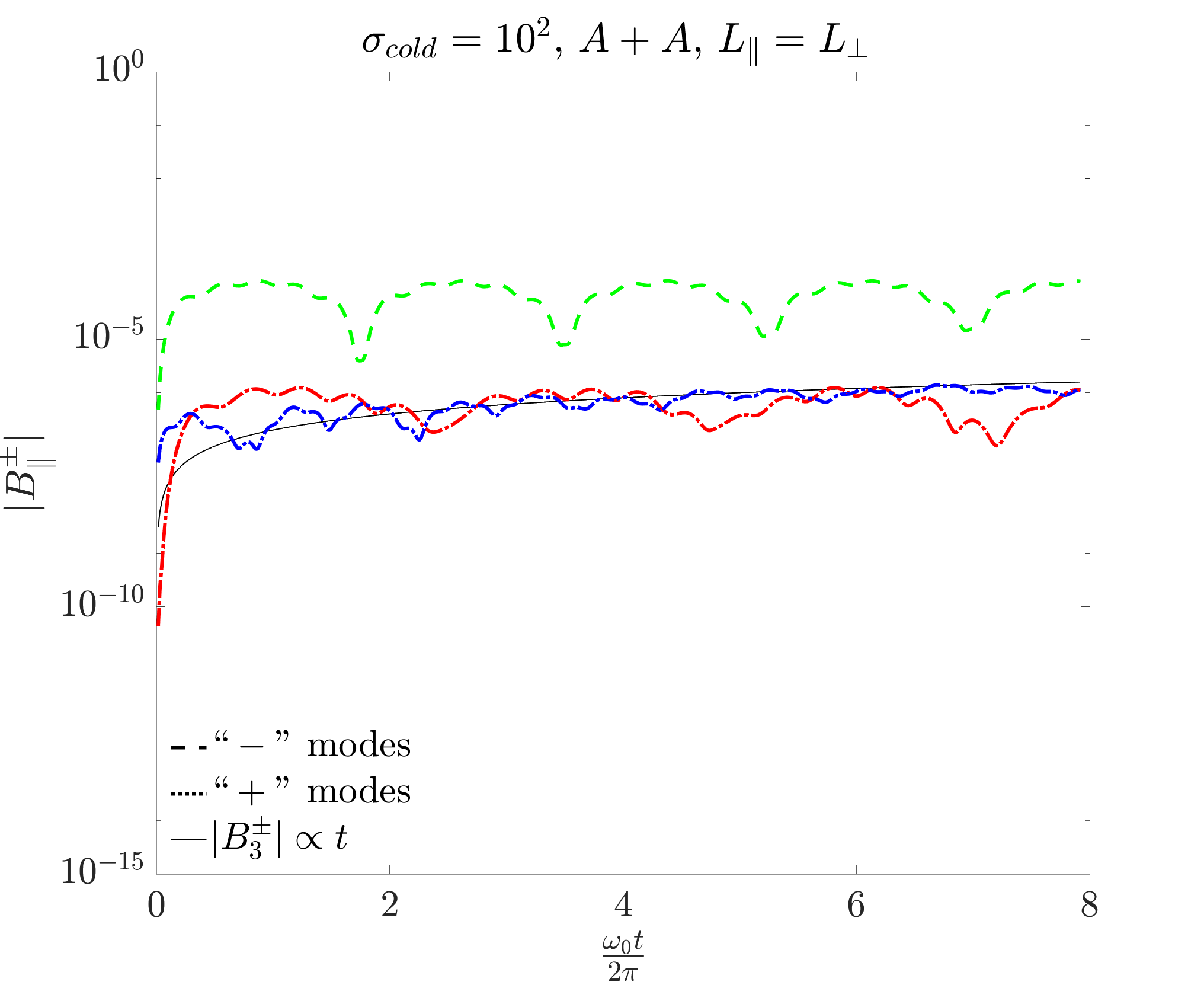}
    
  \includegraphics[width=0.49\textwidth]{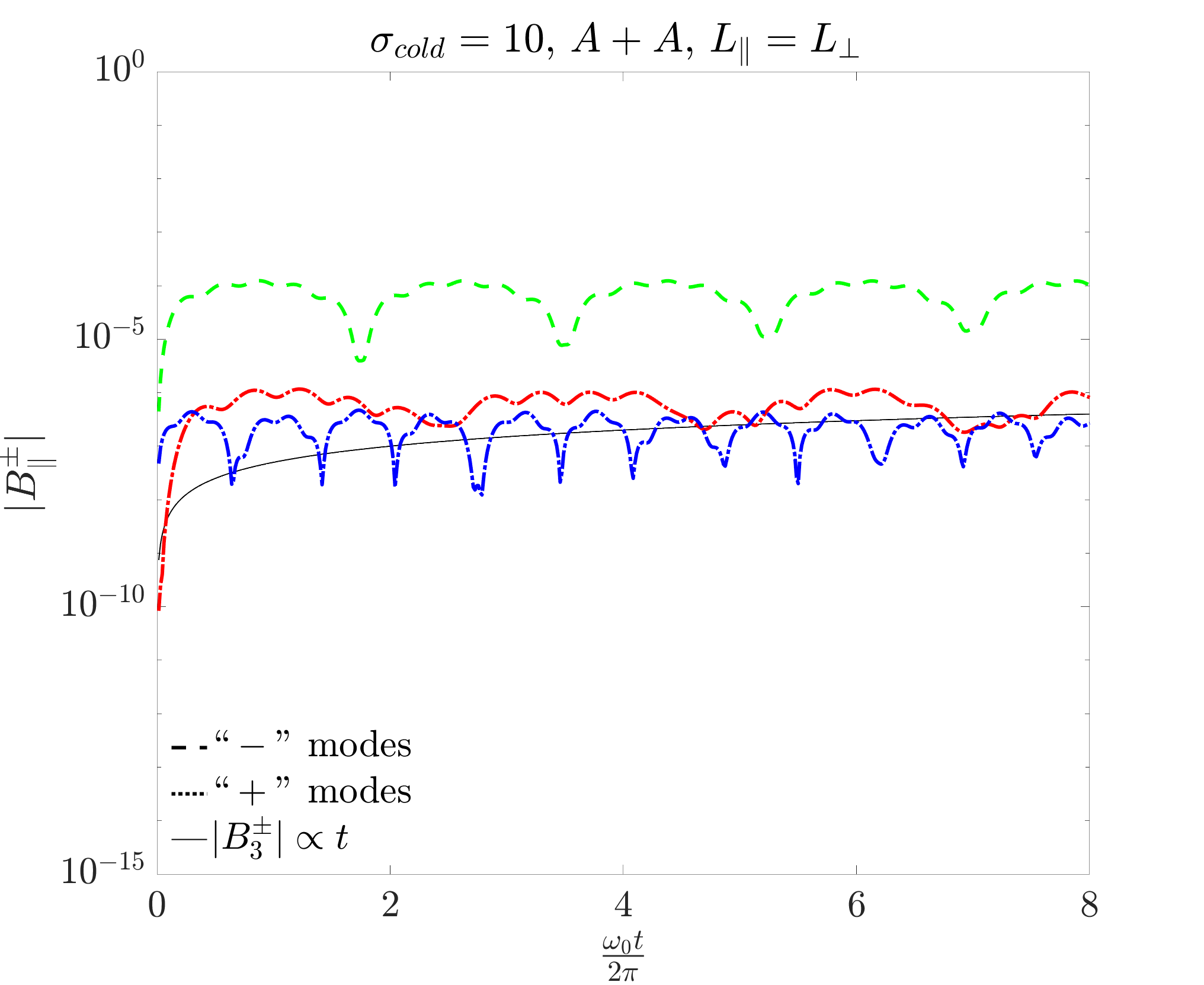}
  \includegraphics[width=0.49\textwidth]{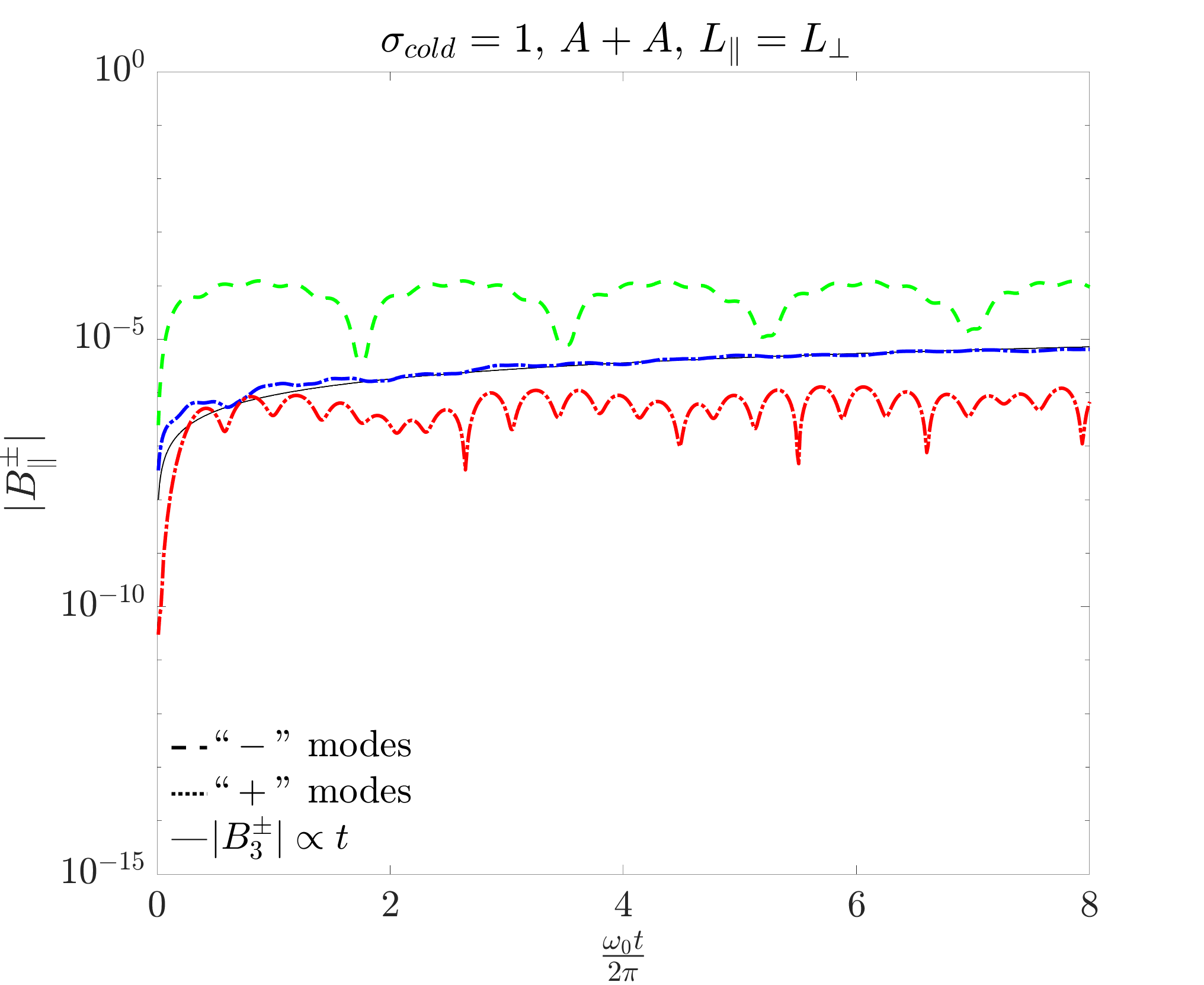}
   
  \includegraphics[width=0.49\textwidth]{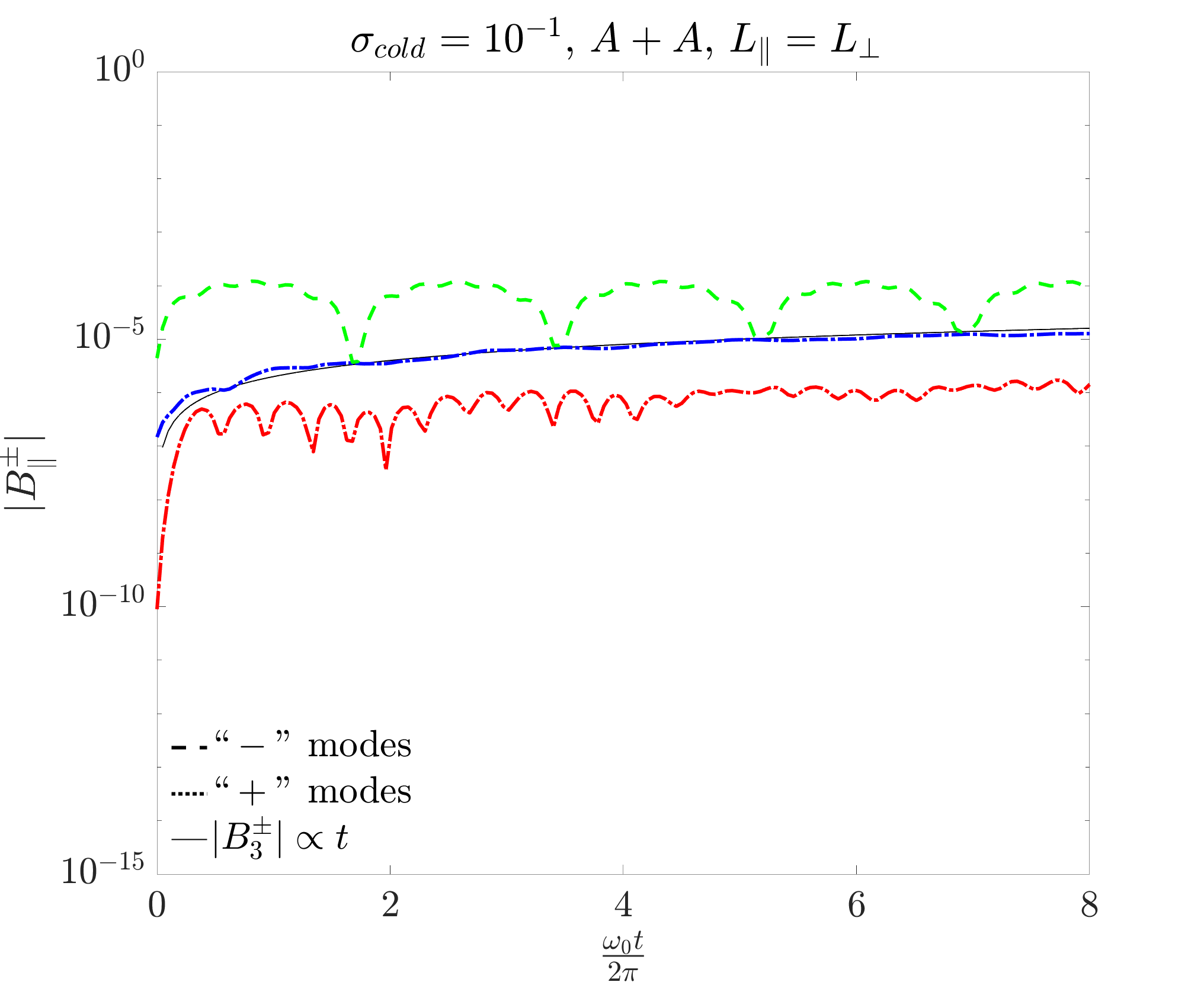}
  \includegraphics[width=0.49\textwidth]{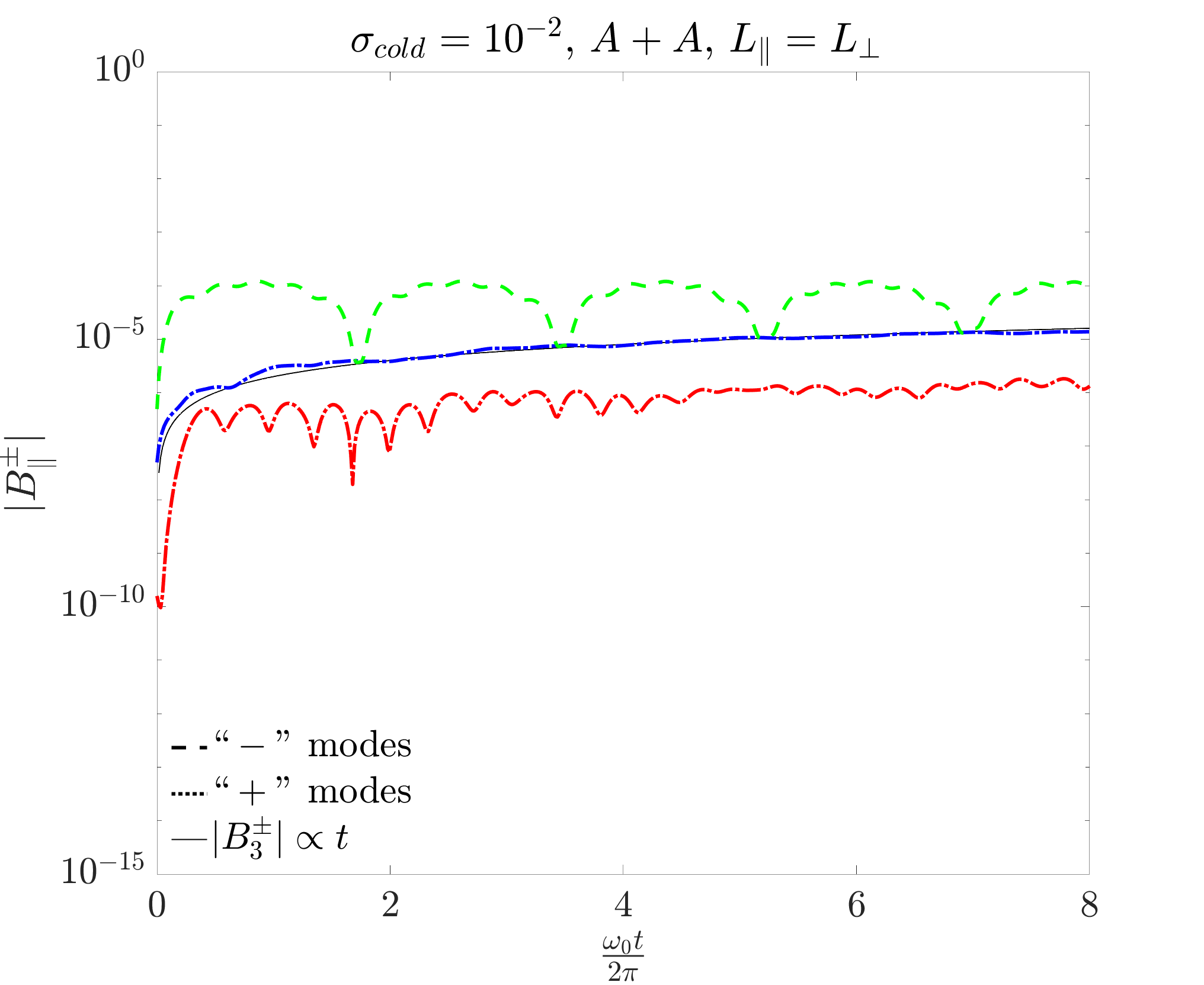}
  \caption{Mode evolution of $B_{\parallel}$ for a cubic domain  with $\chi=0.01$ and $\sigma_{cold} \equiv b^2 / \rho \in [0.01;0.1;1;10;100;\infty]$, presented in descending order of $\sigma$ from top-left to bottom-right. Line styles and colors are as defined in \figref{fig:modesBperp}.}
\label{fig:modesBpar}
\end{figure}

In \figref{fig:modesBpar}, we present the evolution of the highest amplitude $B_\parallel$ modes for the same initial configuration as presented in \figref{fig:modesBperp}. $B_\parallel$ serves as a proxy for the amplitude of compressible, fast mode fluctuations, since the \Alfven\ modes have negligible $B_\parallel$ components. As an additional test (not shown), we have projected the fluctuations onto the fast and \Alfven\ wave eigenfunctions and recovered quantitatively similar results.

We first note that the $(0,1,-1)$ and $(1,0,1)$ $B_\parallel$ modes have zero amplitude at $t=0$ but do develop finite amplitude that is approximately independent of $\sigma$. These modes appear at a mixture of two or more frequencies, the most dominant of which is at the fast mode frequency, $\omega \sim \omega_f \sim k c_f \sim \sqrt{2} \omega_0$, and remain at the same or lower amplitude as the tertiary (blue) modes. The secondary mode, $(1,1,0)$ (green), is the highest amplitude mode and is again composed of two or more frequencies, the most dominant of which are $\omega \sim \omega_f \sim k c_f \sim \sqrt{2} \omega_0$ and $\omega \sim \omega_0 / 2$. Unlike the \Alfven\ branch, the $(1,1,0)$ mode can be a linear fast mode, which is consistent with a component frequency being at the fast mode frequency. Finally, the tertiary modes (blue) again display secular growth with time, with a primary frequency given by the fast mode frequency $\omega \sim \omega_f \sim k c_f \sim \sqrt{6} \omega_0$. The tertiary mode also displays the most significant dependence on $\sigma$. For $\sigma \ll 1$, the amplitude is independent of $\sigma$.

\subsubsection{Elongated Domain}\label{sec:awawelongated}

The scaling of the \Alfven-fast mode coupling discussed above focuses purely on the strength of the background magnetic field while neglecting any wavevector anisotropy that may naturally arise from a strong magnetic field. As noted in \secref{sec:weakComp}, in the Newtonian, obliquely propagating limit ($k_\parallel \lesssim k_\perp$), the fast and \Alfven\ branch turbulent cascades decouple, and we expect an analogous decoupling to occur in the relativistic limit, since the fast and \Alfven\ wave frequencies remain well separated in the oblique limit. Thus, we now examine the results of an elongated box simulation with $L_\parallel = 10L_\perp$.  

\begin{figure}
  \centering
  \includegraphics[width=0.49\textwidth]{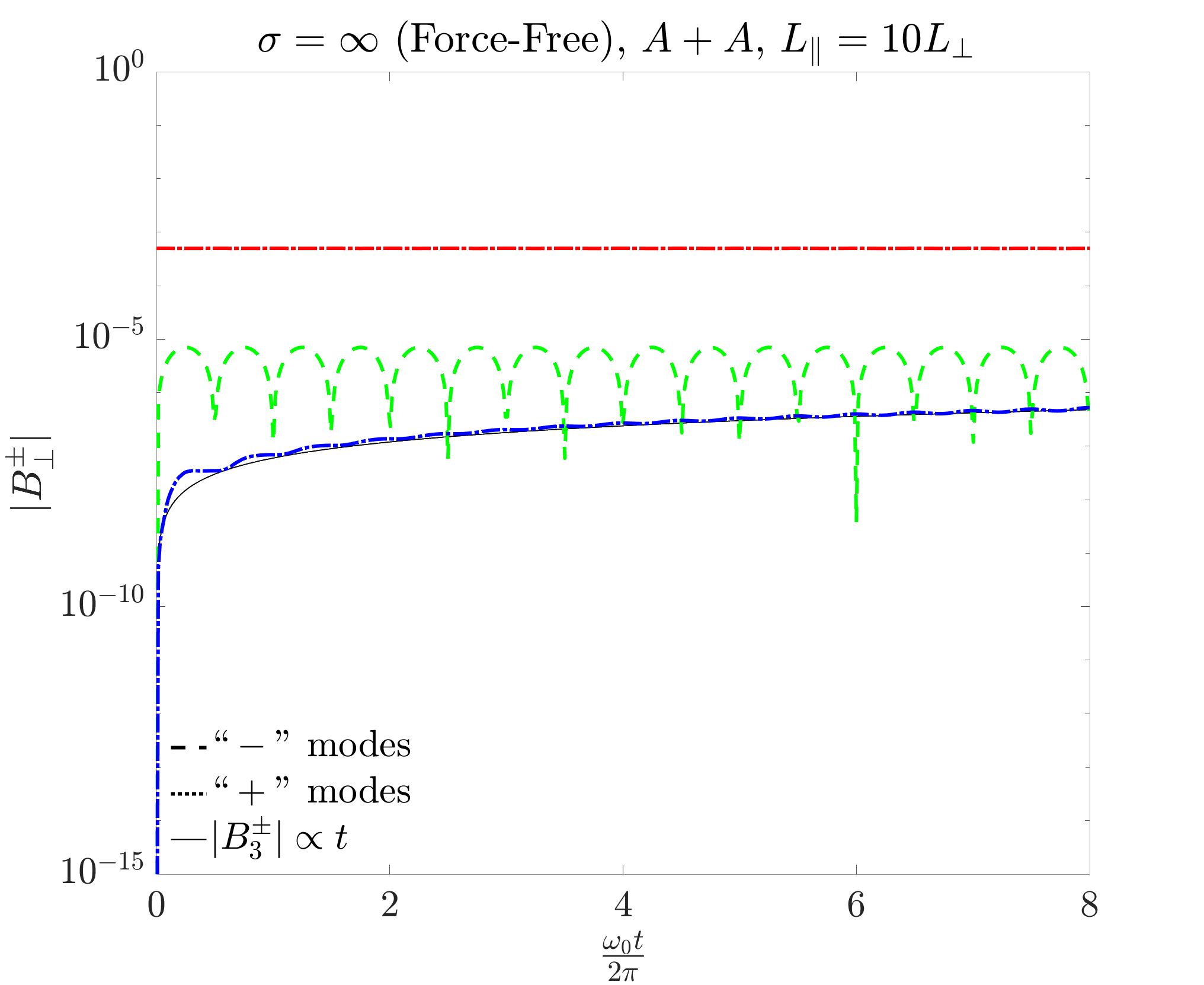}
  \includegraphics[width=0.49\textwidth]{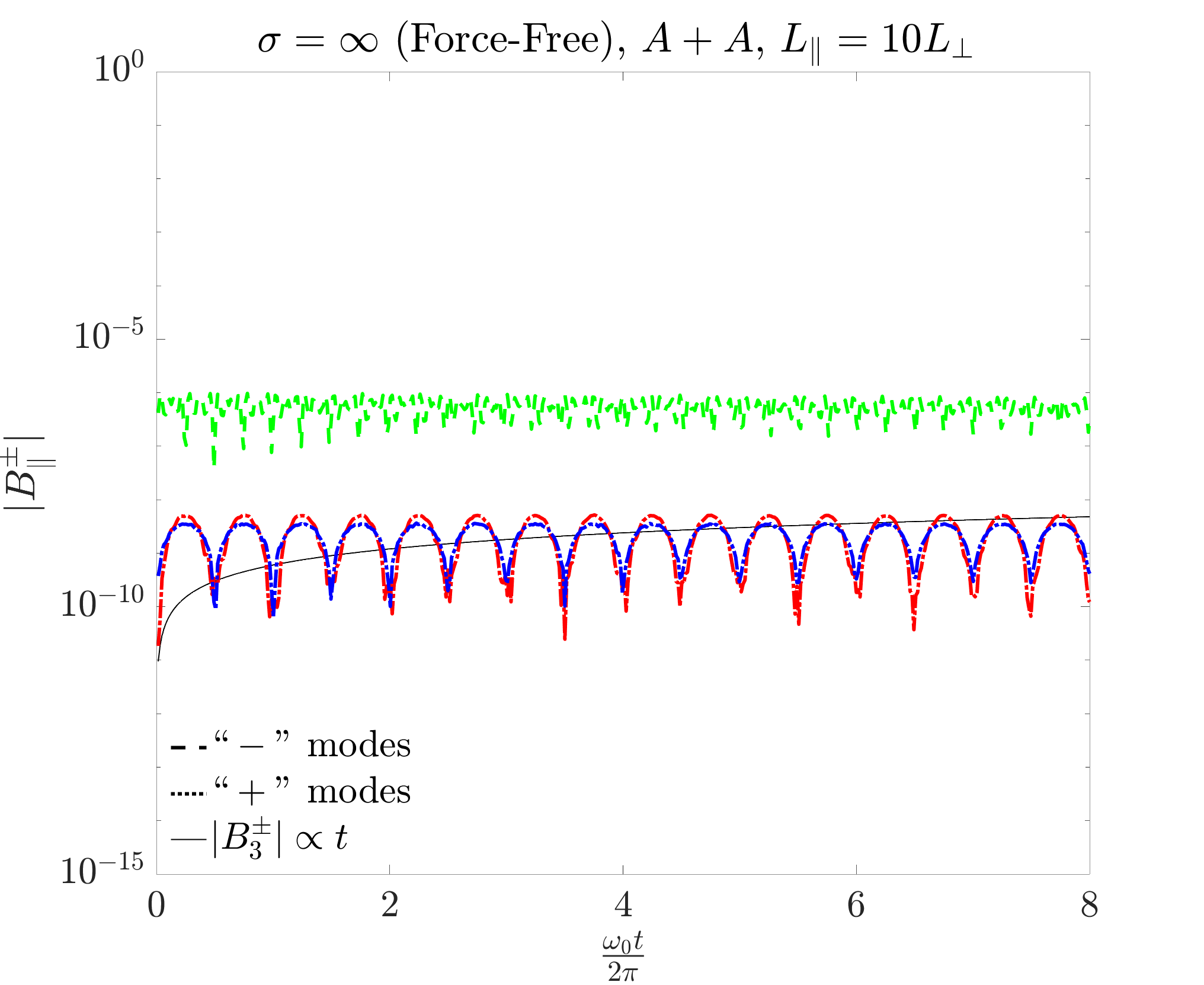}
    
 \includegraphics[width=0.49\textwidth]{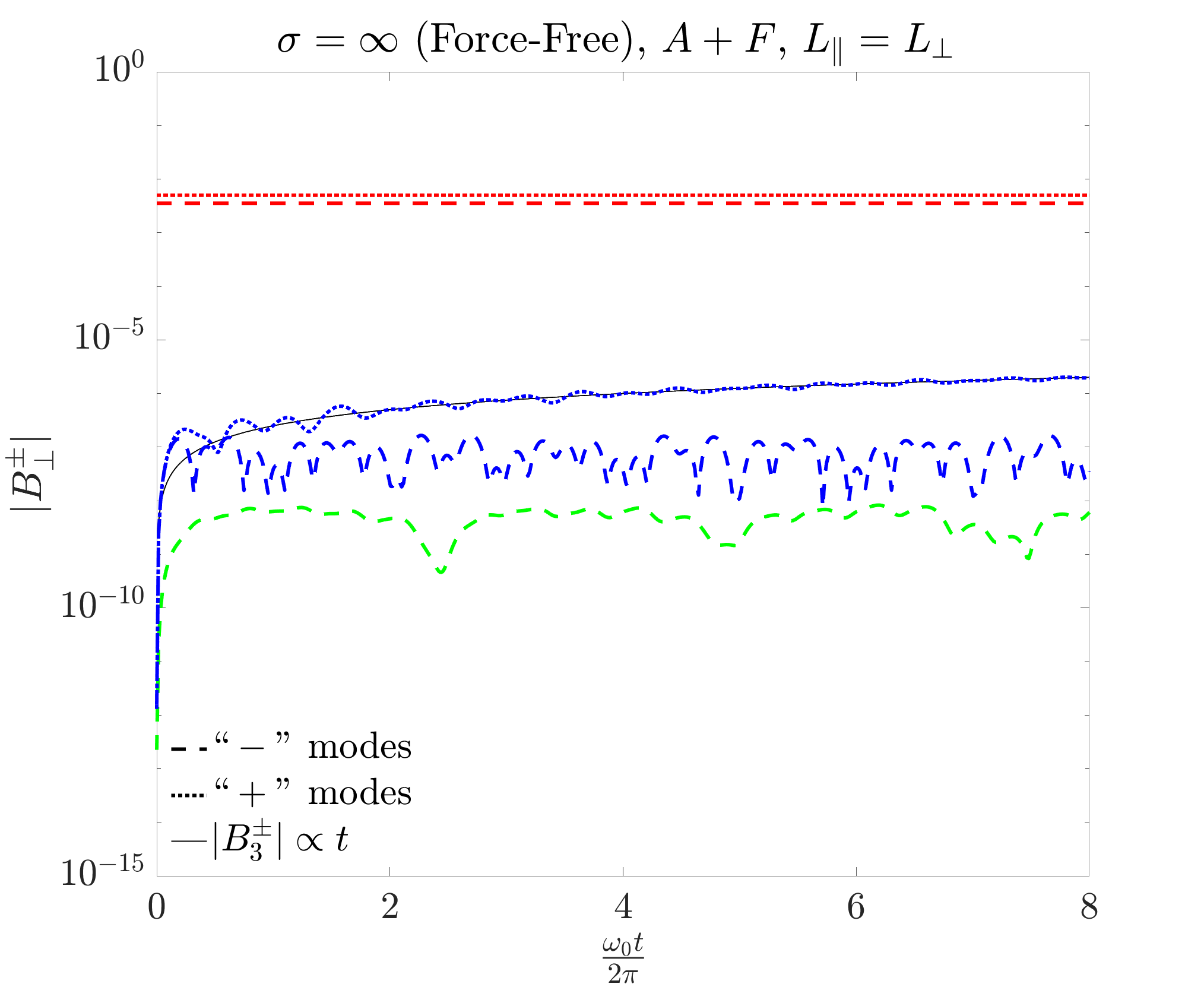}
  \includegraphics[width=0.49\textwidth]{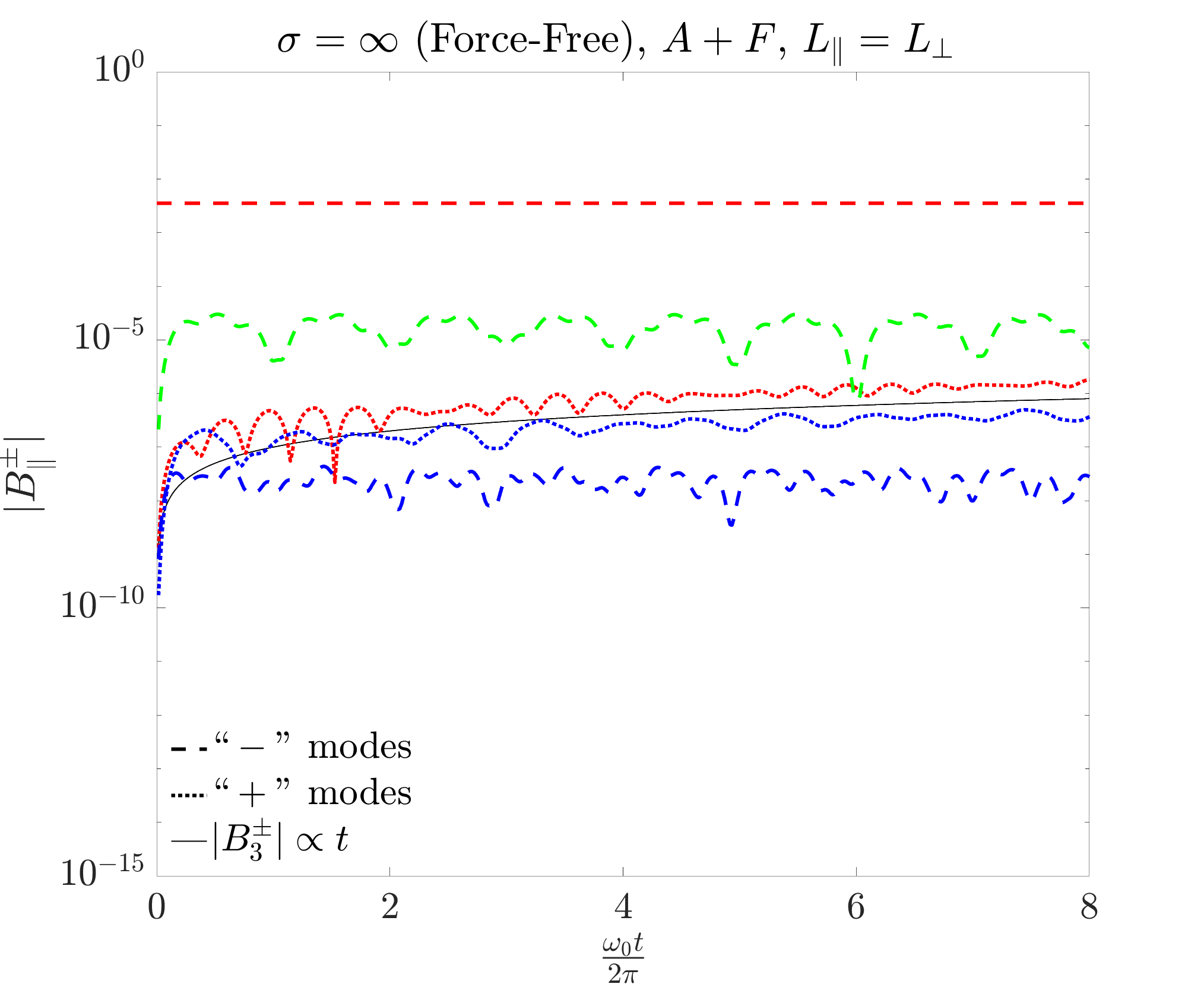}
  
  \includegraphics[width=0.49\textwidth]{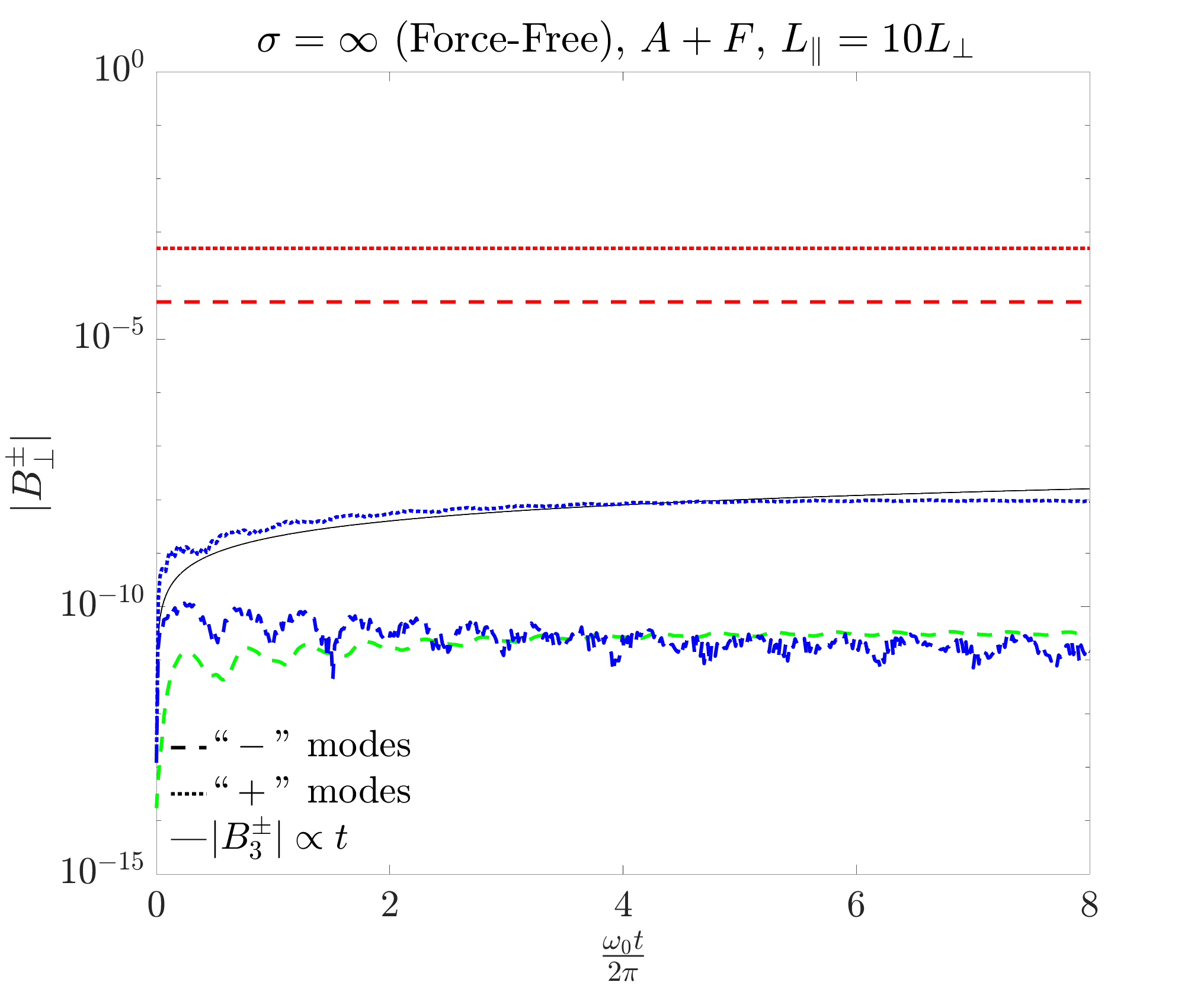}
  \includegraphics[width=0.49\textwidth]{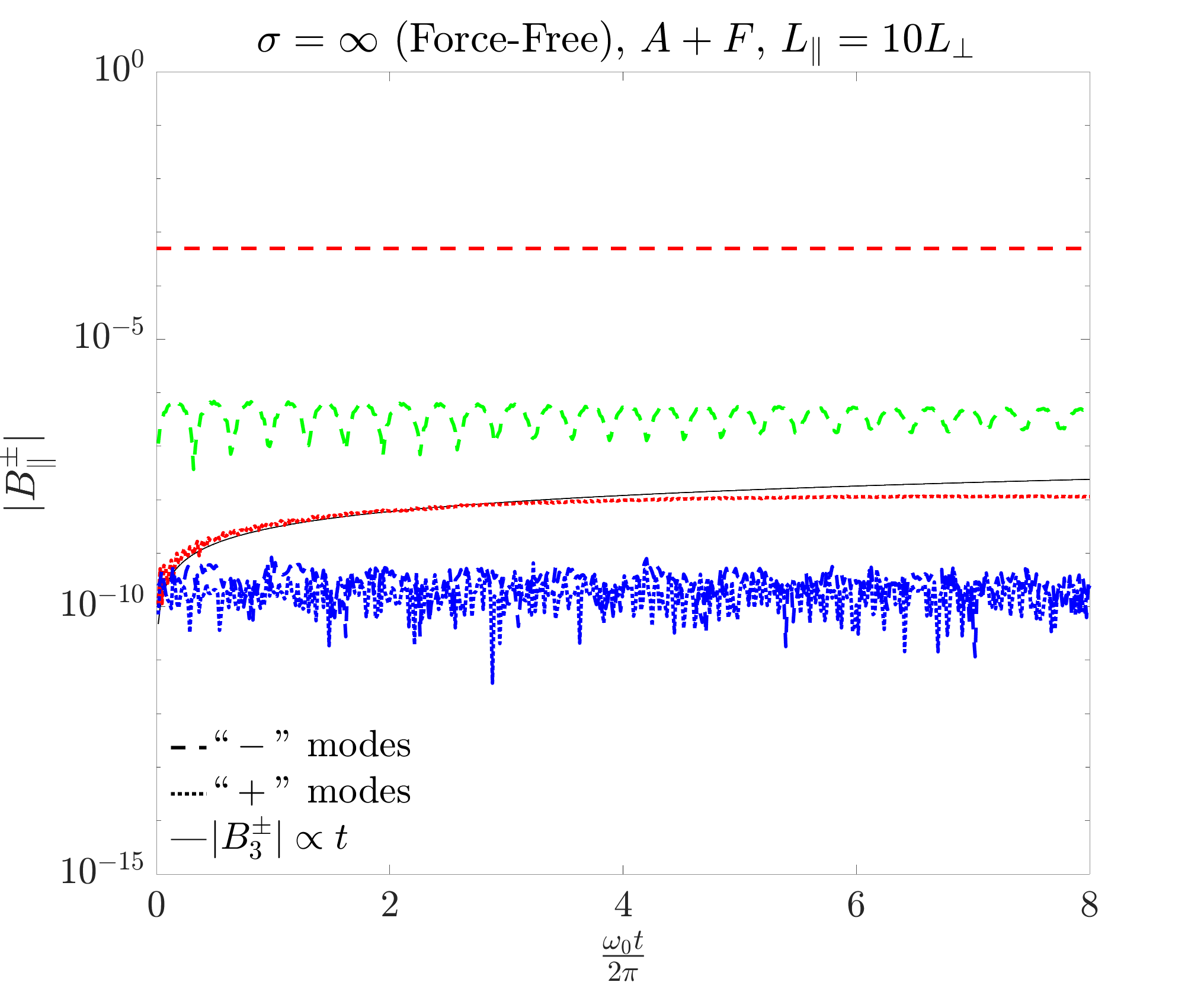}
   
  \caption{Mode evolution of $B_{\perp}$ and $B_{\parallel}$  with $\chi=0.01$. The upper two panels present results for an \Alfven\ wave-\Alfven\ wave collision in an elongated domain, $L_z = 10 L_x$. The middle and lower panels present results for an \Alfven\ wave-fast wave collision in a cubic domain (middle) and elongated domain (lower). The line style and color are as in previous figures.}
\label{fig:modesElongated}
\end{figure}

In the top two panels of \figref{fig:modesElongated} are presented the mode analysis results of a force-free simulation of obliquely propagating \Alfven\ waves, with $k_\perp = 10 k_\parallel$. The color and line styles are as in the previous figures. Comparing the relative amplitudes between each of the $B_\perp$ modes to the relative amplitudes for the cubic domain case presented in the upper left panel of \figref{fig:modesBperp}, it is clear that the elongated domain does not change the \Alfvenic\, $B_\perp$, cascade. However, the three parallel modes' growth, amplitude, and frequency in the elongated case compared to the cubic case in the upper left panel of \figref{fig:modesBpar} are markedly different. First, no secular growth of parallel modes is apparent. If there is a secularly growing mode, it is sufficiently low in amplitude that it does not modulate or interfere with the non-growing, purely oscillatory modes. Second, the mode amplitudes relative to the \Alfven\ wave primaries are an order of magnitude smaller in the elongated case relative to the cubic domain, which is consistent with the compressible mode amplitudes scaling with the elongation factor. Third, in the elongated case, the first (red) and third (blue) order parallel modes are dominated by low frequency oscillations that match the frequency of the \Alfvenic\ secondary mode. The second order (green) mode is much higher in frequency, well above even the fast mode frequency, suggesting nonlinear modes with multiple frequencies contribute to this component. 

These results support our intuition that the \Alfven\ and fast mode cascades decouple as the wavevector anisotropy increases {and extend the findings of \cite{Chandran:2005} to the relativistic limit}. In the case examined here, $\omega_A = k_\parallel \simeq 0.1 \omega_F  = 0.1 k$, leading to a poor frequency matching condition required for the resonant weak turbulence interaction. Although the wave matching condition broadens in strong turbulence, the \Alfven\ and fast mode cascades will remain well separated in the limit $k_\parallel \ll k_\perp$.

\subsection{\Alfven\ Wave-Fast Wave Collisions}\label{sec:fw-aw}
In the \Alfven\ wave-\Alfven\ wave collision case, we have shown that the amplitude of the product fast modes decreases as the initial modes become anisotropic, indirectly demonstrating that the modes decouple. Here, we directly diagnose the coupling of the \Alfven\ and fast mode cascades by examining the collision between an \Alfven\ wave and a fast wave in both cubic and elongated domains.  Unlike the \Alfven\ wave-\Alfven\ wave collision explored in the previous sections, there is not a canonical choice for the counter-propagating fast mode. Therefore, we choose a fiducial fast mode with $\Vk = (0,1,-1)$, and an \Alfven\ wave with  $\Vk = (1,0,1)$. This choice is largely for consistency with the \Alfven\ wave-\Alfven\ wave collision; however, it also obeys reasonable wavevector and frequency matching conditions, and it has the same expected product modes as the \Alfven\ wave-\Alfven\ wave collision. Further, the secondary modes will still be $(1,1,0)$ and $(-1,1,-2)$, and for these initial modes in a cubic domain, the secondary $(1,1,0)$ mode can satisfy the linear fast mode dispersion relation. For these simulations, we will continue the convention of fixing $\chi = 0.01$ to satisfy weak turbulence.

\subsubsection{Cubic Domain}\label{sec:awfwcubic}
In the middle two panels of \figref{fig:modesElongated} are plotted the results of the \Alfven\ wave-fast wave collision outlined above for a cubic domain. Note that in this case, the dotted and dashed red lines in the $B_\perp$ panel correspond to the initial \Alfven\ and fast waves respectively. Focusing on the behavior in the middle left $B_\perp$ panel, the secondary, green, mode is the lowest amplitude mode, displays no apparent secular growth, and contains a mixture of linear and nonlinear frequencies. Similarly, the minus tertiary is oscillatory and dominated by high frequency nonlinear modes. However, the plus tertiary mode does exhibit secular growth, and is dominated by the linear \Alfven\ wave frequency. Relative to the \Alfven\ wave-\Alfven\ wave collision in the upper left panel of \figref{fig:modesBperp}, this tertiary mode is decreased in amplitude by approximately a factor of three. 

Turning to the middle right $B_\parallel$ panel for the \Alfven\ wave-fast wave in \figref{fig:modesElongated}, we see different behavior compared to the upper left panel of \figref{fig:modesBpar}. The largest amplitude component remains the secondary mode, which is dominated by the fast mode linear frequency and an additional low frequency component. Both the plus mode ``primary'' and tertiary exhibit secular growth with comparable amplitude. Note that the plus mode $B_\parallel$ ``primary'' is not initialized since the initial plus mode is an \Alfven\ wave. This ``primary'' mode growth is due to the interaction of the $(-1,1,2)$ secondary with the fast mode primary with $(0,1,-1)$ summing to give $(1,0,1)$, i.e., the ``primary'' is here a tertiary mode.

\subsubsection{Elongated Domain}\label{sec:awfwelongated}
In the bottom two panels of \figref{fig:modesElongated} are plotted the results of the \Alfven\ wave-fast wave collision outlined above for the elongated domain with $L_\parallel = 10L_\perp$. The overall behavior is similar to the \Alfven\ wave-fast wave collision in the cubic domain with a few notable differences. Focusing first on the $B_\perp$ panel in the lower left of \figref{fig:modesElongated}, we see that the \Alfvenic\ tertiary is further suppressed in amplitude, now reduced by nearly two orders of magnitude relative to the cubic and elongated domain \Alfven\ wave-\Alfven\ wave collisions in the upper left panels of \figref{fig:modesBperp} and \figref{fig:modesElongated}. The secondary and minus tertiary are also further suppressed in amplitude, and the secondary is dominated by higher frequency modes relative to the cubic domain. Turning to the lower right $B_\parallel$ panel for the \Alfven\ wave-fast wave in \figref{fig:modesElongated}, we again see that all product modes are suppressed by approximately an order of magnitude, consistent with the elongation factor weakening the interaction. Notably, both blue tertiary modes display no secular growth. The secondary mode is higher frequency, and the ``primary'' mode is the only $B_\parallel$ component with clear secular growth; however, the ``primary'' is now dominated by a high frequency, nonlinear component. These results further support our intuition {from Newtonian weak turbulence \citep{Chandran:2005}} that the \Alfven\ and fast mode cascades decouple as the wavevector anisotropy increases by directly comparing the \Alfven\ wave-fast wave collision in both cubic and elongated domains extend the findings of  to the relativistic limit.

\section{Summary \& Conclusions}\label{sec:conclusions}

In this paper, we present an analytical and numerical study of relativistic, weak \Alfvenic\ turbulence {focusing on the three-wave interaction}. We begin by reviewing the knowledge gained from  non-relativistic (Newtonian) turbulence and use that framework to guide our study of relativistic turbulence. {From Newtonian turbulence theories,  we know that both weak and strong \Alfvenic\ turbulence lead to anisotropic cascades.  Therefore, regardless of the strength of the turbulence and isotropy at large scales, turbulence will become anisotropic and eventually satisfy $k_\parallel \ll k_\perp$. Leveraging the anisotropic cascade of energy and the fact that $\delta B/B_0 \ll 1$ is satisfied either at the outer-scale or once the cascade reaches sufficiently small scales,  we derive a set of relativistic reduced MHD (RRMHD) equations and cast them in Elsasser form. This set of equations has the same properties and basic form as RMHD and is appropriate for describing strongly magnetized relativistic plasmas for which \Alfvenic\ fluctuations dominate. We note that RRMHD is not equivalent to the magnetically dominated equations of the force-free limit of relativistic MHD, which require $\sigma = b^2 / h \rightarrow \infty$.  However, we do also derive a set of magnetically dominated RRMHD equations which are valid in the $\sigma \rightarrow \infty$ limit but retain the favorable properties inherent to RMHD for the analysis of \Alfvenic\ turbulence.}

{We note that the similarity between RMHD and RRMHD extends to the wave kinetic equation, which describes the spectral evolution  of a weakly turbulence plasma. As such, we conclude that RRMHD: (i) is dominated by three-wave interactions of \Alfven\ waves; and (ii) leads to an energy spectrum of the form $f(k_\parallel) k_\perp^{-2}$, where $f(k_\parallel)$ is set by external forcing or initial conditions. We then employ the RRMHD equations in Elsasser form to analytically compute the building blocks of weak turbulence through third order to heuristically demonstrate the primacy of the three-wave interaction.} The primary interaction is the collision of perpendicularly polarized \Alfven\ waves with $\omega = \omega_0$ and wavevectors $\Vk_{1\pm} = (1,0,1)$  and $(0,1,-1)$ that interact through a three-wave interaction to produce a non-linear, magnetic shear mode, $\Vk_2 = (1,1,0)$, with $\omega = 2\omega_0$ at second order. The interaction of this secondary mode with the primaries through a subsequent three-wave interaction produces at third order two secularly growing linear \Alfven\ waves with $\Vk_3 = (1,2,1)$ and $(2,1,-1)$, confirming the prediction that there is no parallel cascade of energy. The analytical results also confirm the constraint that upward and downward propagating fluctuations do not exchange energy. In other words, the upward propagating primary mode transfers energy to the upward propagating tertiary mode, and the downward propagating primary mode transfers energy to the downward propagating tertiary. These analytical solutions are fundamentally identical to the Newtonian results of \cite{Howes:2013a} and highlight the fundamental role three-wave interactions of \Alfven\ waves play in transferring energy from large to small scales in relativistic plasmas.

{Since there is not a formally incompressible limit in relativistic systems wherein there is a finite speed of propagation, $c$, we turn to numerical simulations to confirm these analytical results and test the coupling of the \Alfven\ and fast modes in relativistic weak turbulence. We} perform a set of numerical simulations for $\sigma \in [0.01;0.1;1;7;20;\infty]$ while keeping $\chi = 0.01$ fixed to maintain weak  turbulence. We examine both \Alfven\ wave-\Alfven\ wave collisions as well as \Alfven\ wave-fast wave collisions in  both cubic and elongated domains, where the elongated domain is employed to approximate the reduced limit of $k_\parallel \ll k_\perp $. The numerical results confirm the analytical findings for the case of \Alfven\ wave-\Alfven\ wave collisions in both cubic and elongated domains, and the \Alfvenic\ cascade is unaltered by the elongated domain. We find that in the cubic domain, both the \Alfven\ wave-\Alfven\ wave  and \Alfven\ wave-fast wave collisions produce secondary and tertiary fluctuations consistent with fast waves. However, in the elongated domains, the \Alfven\ wave-fast wave interaction is suppressed, and the suppression is proportional to the elongation factor, thus confirming our theoretical expectation that the two modes decouple in the anisotropic limit. More detailed numerical analysis of relativistic \Alfven\ wave and \Alfven\ wave packet collisions, including the formation and evolution of current sheets, can be found in Paper II, \citep{Ripperda:2021}.

The results of this analytical and numerical study of weak, relativistic, \Alfvenic\ turbulence present a simple picture of nonlinear energy transfer through \Alfven\ wave collisions by highlighting the importance of the three-wave interaction and the decoupling of the fast and \Alfven\ modes in the anisotropic, reduced limit. Further, they demonstrate the fundamental importance of \Alfvenic\ turbulence to high energy astrophysical systems, extending the work of {\cite{Ng:1996,Galtier:2000,Chandran:2005,Howes:2013a}}. With these insights on the continued importance of \Alfvenic\ interactions even in the relativistic limit, we may examine turbulence in these extreme astrophysical systems with newfound intuition.


\section*{Acknowledgements} 
We thank Gregory Howes and Ben Chandran for helpful discussions. We acknowledge the Flatiron's Center for Computational Astrophysics (CCA) and the Princeton Plasma Physics Laboratory (PPPL) for support of collaborative CCA-PPPL meetings on plasma-astrophysics where the ideas presented in this paper have been initiated. The computational resources and services used in this work were provided by facilities supported by the Scientific Computing Core at the Flatiron Institute, a division of the Simons Foundation; And by the VSC (Flemish Supercomputer Center), funded by the Research Foundation Flanders (FWO) and the Flemish Government – department EWI. This work was supported by the Simons Foundation (JMT, AB, and AC), (YY, Flatiron Research Fellowship); and a Joint Princeton/Flatiron Postdoctoral Fellowship (BR); the National Science Foundation (AP and JFM, Grant No. AST-1909458), (JJ, Atmospheric and Geospace Science Postdoctoral Fellowship, Grant No. AGS-2019828);  NSF Atmospheric and Geospace Science Postdoctoral Fellowship (JJ, Grant No. AGS-2019828); a joint fellowship at the Princeton Center for Theoretical Science, the Princeton Gravity Initiative, and the Institute for Advanced Study (ERM). Declaration of Interests: The authors report no conflict of interest.


\appendix
\section{Ordered Equations}\label{sec:appA}
\subsection{Conversion to Elsasser Potentials and Characteristic Variables}
To prepare the system of equations for analysis of their asymptotic solutions, we will convert to the Elsasser potential formulation.  This approach eliminates the nonlinear pressure term, and reduces the equation system to two scalar equations for the Elsasser potentials, $\zeta_\pm$ rather than two vector equations for the Elsasser fields, $\Vz_\pm$. \footnote{For the remainder of the analysis, we will focus on the three-vector form of the relativistic, reduced Elsasser equations rather than the covariant form. This is done for simplicity and to avoid unnecessary covariant projection four-vectors and curl formulations.} The Elsasser potentials are defined by the relation $\delta \Vz_\pm = \hz \times \grad_\perp \zeta_\pm$, which follows from the fact that the Elsasser fields are solenoidal. Note that $\delta \Vv_\perp$ and $\delta \VB_\perp$ can be reconstructed from the potentials
\begin{equation}
    \begin{split}
        \delta \Vv_\perp &= \frac{1}{2} \hz \times \grad_\perp (\zeta_+ + \zeta_-)\\
        \frac{\delta \VB_\perp}{\sqrt{\mathcal{E}_0}} &= \frac{1}{2} \hz \times \grad_\perp (\zeta_+ - \zeta_-).
    \end{split}
\end{equation}
Also, using the fact that in general in relativistic MHD we use $\V{E} = - \Vv \times \VB$, to lowest order
\begin{equation}
 \frac{\V{E}_\perp}{B_0} = - \frac{1}{2} \grad_\perp (\zeta_+ +\zeta_-).
\end{equation}

Taking  the curl of \eqr{eq:RRElsasser3} and substituting the expression for the Elsasser potentials yields the Elasser potential equations 
\begin{equation}\label{eq:RRPElsasser}
    \pfrac{\grad_\perp^2 \zeta_\pm}{t} \mp v_A \pfrac{\grad_\perp^2 \zeta_\pm}{z} = -\frac{1}{2}\left[ \left\{\zeta_+,\grad_\perp^2 \zeta_-\right\} + \left\{\zeta_-,\grad_\perp^2 \zeta_+\right\} \mp \grad_\perp^2 \left\{ \zeta_+, \zeta_-\right\}\right],
\end{equation}
which are identical in form to those derived by \citep{Schekochihin:2009}. The Poisson bracket is defined by
\begin{equation}
    \left\{f,g\right\} = \hz \cdot (\grad_\perp f \times \grad_\perp g).
\end{equation}

\eqr{eq:RRPElsasser} retains the form that the left-hand side describes the linear evolution, and we can further simplify the task of solving the equation set by converting to characteristic variables, $\phi_\pm = z \pm v_A t$\footnote{Note that this variable transformation remains valid even in the force-free limit in which $v_A = c$, because the frame remains inertial.}. In terms of these characteristic variables, \eqr{eq:RRPElsasser} becomes
\begin{equation}\label{eq:RRPCElsasser}
    \pfrac{\grad_\perp^2 \zeta_\pm}{\phi_\mp}  = \pm \frac{1}{4 v_A}\left[ \left\{\zeta_{\alert{+}},\grad_\perp^2 \zeta_{\alert{-}}\right\} + \left\{\zeta_{\alert{-}},\grad_\perp^2 \zeta_{\alert{+}}\right\} \mp \grad_\perp^2 \left\{ \zeta_{\alert{+}}, \zeta_{\alert{-}}\right\}\right].
\end{equation}

Finally, the Elsasser potential form for the initial conditions provided in ~\eqr{eq:ICs} is
\begin{equation}
    \begin{split}
        \zeta_{1+} &= \quad\!\frac{z_+}{k_\perp} \sin{(k_\perp x - k_\parallel z - \omega_0 t)} = \quad\!\frac{z_+}{k_\perp} \sin{(k_\perp x - k_\parallel \phi_+)}\\
        \zeta_{1-} &= -\frac{z_-}{k_\perp} \sin{(k_\perp y + k_\parallel z - \omega_0 t)} = -\frac{z_-}{k_\perp} \sin{(k_\perp y + k_\parallel \phi_-)},
    \end{split}
\end{equation}
where the final equality follows from conversion to the characteristic variables $\phi_\pm$.

\subsection{Linear, $\order{\varepsilon}$, Solutions}\label{sec:order1}
At lowest, linear, order, \eqr{eq:RRPCElsasser} reduces to
\begin{equation}\label{eq:RRPCLElsasser}
    \pfrac{\grad_\perp^2 \zeta_{1\pm}}{\phi_\mp} = 0.
\end{equation}
The initial conditions above satisfy \eqr{eq:RRPCLElsasser} if $\omega_0$ is the linear \Alfven\ wave frequency, $\omega_0 = k_\parallel v_A$. Thus, at lowest order the solution describes counter-propagating, linear \Alfven\ waves, as expected of RRMHD. 

\subsection{Secondary, $\order{\delta^2}$ Solutions}\label{sec:order2}
At second order, the evolution  equations become
\begin{equation}\label{eq:RRPC2Elsasser}
    \pfrac{\grad_\perp^2 \zeta_{2\pm}}{\phi_\mp}  = \pm \frac{1}{4 v_A}\left[ \left\{\zeta_{1{\alert{+}}},\grad_\perp^2 \zeta_{1{\alert{-}}}\right\} + \left\{\zeta_{1{\alert{-}}},\grad_\perp^2 \zeta_{1{\alert{+}}}\right\} \mp \grad_\perp^2 \left\{ \zeta_{1{\alert{+}}}, \zeta_{1{\alert{-}}}\right\}\right],
\end{equation}
and we can simply insert our $\zeta_{1\pm}$ solutions to solve for $\zeta_{2\pm}$. Upon substitution, we note that the first two nonlinear terms on the left-hand side cancel, leaving
\begin{equation}
\begin{split}
\frac{\partial \grad_\perp^2 \zeta_{2\pm}}{\partial \phi_\mp}
= \frac{-k_\perp^2 z_+z_-}{4v_A} \left\{ 
\cos\left[k_\perp x+k_\perp y- k_\parallel (\phi_+ - \phi_-)\right] \right.
+   \\ \left. \cos\left[k_\perp x - k_\perp y - k_\parallel (\phi_+ + \phi_-)\right] 
 \right\}.
 \end{split}
\end{equation}
Integrating the above equations from $t'=0$ to $t' = t$, i.e., from $\phi'_\pm = \phi_+ = \phi_-$ to $\phi_+' = \phi_+$ and $\phi_-' = \phi_-$, yields the $\order{\varepsilon^2}$ solutions
\begin{eqnarray}\label{eq:zeta2+phi}
\zeta_{2+} & = \frac{ z_+ z_-}{8 \omega_0} &\left\{ 
\sin[k_\perp x + k_\perp y - k_\parallel (\phi_+-\phi_-)] - \sin[k_\perp x + k_\perp y ]  \right. 
\nonumber \\
& &- \left.  \sin[k_\perp x - k_\perp y - k_\parallel (\phi_+ + \phi_-) ] 
+\sin[k_\perp x - k_\perp y - 2 k_\parallel \phi_+ ] \right\},
\end{eqnarray}

\begin{eqnarray}\label{eq:zeta2-phi}
\zeta_{2-} & = -\frac{ z_+ z_-}{8 \omega_0} &\left\{ 
\sin[k_\perp x + k_\perp y - k_\parallel (\phi_+-\phi_-)] - \sin[k_\perp x + k_\perp y ]  \right. 
\nonumber \\
& &+  \left.  \sin[k_\perp x - k_\perp y - k_\parallel (\phi_+ + \phi_-) ] 
-\sin[k_\perp x - k_\perp y - 2 k_\parallel \phi_- ] \right\},
\end{eqnarray}
or in terms of $z$ and $t$
\begin{eqnarray}\label{eq:zeta2+}
\zeta_{2+} & = \frac{ z_+ z_-}{8 \omega_0} &\left\{ 
\sin[k_\perp x + k_\perp y - 2 \omega_0 t ] - 
\sin[k_\perp x + k_\perp y ]  \right. \nonumber \\
& &-  \left.  \sin[k_\perp x - k_\perp y - 2 k_\parallel z ]
+\sin[k_\perp x - k_\perp y - 2 k_\parallel z - 2 \omega_0 t ] \right\},
\end{eqnarray}

\begin{eqnarray}\label{eq:zeta2-}
\zeta_{2-} & = -\frac{ z_+ z_-}{8 \omega_0} &\left\{ 
\sin[k_\perp x + k_\perp y - 2 \omega_0 t ] - \sin[k_\perp x + k_\perp y ]  \right. \nonumber \\
& &+  \left.  \sin[k_\perp x - k_\perp y - 2 k_\parallel z] 
-\sin[k_\perp x - k_\perp y - 2 k_\parallel z + 2 \omega_0 t  ] \right\}.
\end{eqnarray}
Converting the second order Elsasser potential solutions into solutions for $\VB_{\perp2}$ and $\mathbf{E}_{\perp2}$
\begin{equation}\label{eq:O2B}
\begin{split}
\frac{\V{B}_{\perp 2}}{\sqrt{\mathcal{E}_0}} &= \frac{  z_+ z_-}{16 v_A} 
\frac{  k_\perp }{k_\parallel}
\left\{
\left[ 2 \cos(k_\perp x + k_\perp y - 2 \omega_0 t ) - 
2 \cos(k_\perp x + k_\perp y )  \right] (-\xhat+ \yhat) \right.  \\
 &+  \left. \left[ \cos(-k_\perp x + k_\perp y +2 k_\parallel z + 2 \omega_0 t )  -\cos(- k_\perp x + k_\perp y + 2 k_\parallel z - 2 \omega_0 t ) \right](\xhat+ \yhat) \right\},
\end{split}
\end{equation}
\begin{eqnarray}\label{eq:O2E}
\frac{\V{E}_{\perp 2}}{B_0} & = -\frac{  z_+ z_-}{16 v_A} 
\frac{  k_\perp }{k_\parallel}
&\left\{ \left[ 2 \cos(-k_\perp x + k_\perp y +2 k_\parallel z ) -
\cos(-k_\perp x + k_\perp y +2 k_\parallel z + 2 \omega_0 t ) 
 \right. \right. \nonumber \\
& &  \left. \left. - \cos(- k_\perp x + k_\perp y + 2 k_\parallel z - 2 \omega_0 t ) 
\right](-\xhat+ \yhat)  \right\}.
\end{eqnarray}

\subsection{Tertiary, $\order{\varepsilon^3}$ Solutions}\label{sec:order3}
At third order, the evolution equations for the Elsasser potentials become
\begin{equation}\label{eq:RRPC3Elsasser}
\begin{split}
\frac{\partial  \nabla_\perp^2 \zeta_{3\pm}}{\partial \phi_\mp}
= \pm \frac{1}{4v_A} &\left[ \{\zeta_{1{\alert{+}}},  \nabla_\perp^2 \zeta_{2{\alert{-}}}\}
+  \{\zeta_{2{\alert{+}}},  \nabla_\perp^2 \zeta_{1{\alert{-}}}\}
+  \{\zeta_{1{\alert{-}}},  \nabla_\perp^2 \zeta_{2{\alert{+}}}\} \right.\\
&\left.+  \{\zeta_{2{\alert{-}}},  \nabla_\perp^2 \zeta_{1{\alert{+}}}\} 
\mp \nabla_\perp^2 \{\zeta_{1{\alert{+}}},   \zeta_{2{\alert{-}}}\} 
\mp \nabla_\perp^2 \{\zeta_{2{\alert{+}}},   \zeta_{1{\alert{-}}}\} \right].
\end{split}
\end{equation}
Substituting the lower order solutions into the nonlinear terms,
\begin{eqnarray}\label{eq:eq3plus}
\frac{\partial \nabla_\perp^2 \zeta_{3+}}{\partial \phi_-} & = \frac{ z_+^2 z_- k_\perp^3}{64 \omega_0 v_A} &\left\{ 
4 \cos[2 k_\perp x + k_\perp y - k_\parallel \phi_+] - 4 \cos[2 k_\perp x + k_\perp y - 2 k_\parallel \phi_+ + k_\parallel \phi_-]  \right. \nonumber\\
& & + 4 \cos[-2 k_\perp x + k_\perp y + 2 k_\parallel \phi_+ +  k_\parallel \phi_-] \nonumber\\
&&\left.-4 \cos[-2 k_\perp x + k_\perp y + k_\parallel \phi_+ + 2 k_\parallel \phi_-] \right\} \\
& + \frac{ z_+ z_-^2 k_\perp^3}{64 \omega_0 v_A} &\left\{ 
6 \cos[ k_\perp x + 2 k_\perp y  + k_\parallel \phi_-] 
- 6 \cos[ k_\perp x + 2 k_\perp y - k_\parallel \phi_+ + 2 k_\parallel \phi_-]  \right. \nonumber\\
& &  +6 \cos[ -k_\perp x + 2 k_\perp y + k_\parallel \phi_+ + 2 k_\parallel \phi_-] \nonumber\\
 &&-6 \cos[ -k_\perp x + 2 k_\perp y +2 k_\parallel \phi_+ + k_\parallel \phi_-] 
 \nonumber\\
&&  \left. + 2  \cos[ k_\perp x - k_\parallel \phi_-] 
- 2 \cos[ k_\perp x -2  k_\parallel \phi_+ + k_\parallel \phi_-] 
 \right\},\nonumber 
\end{eqnarray}
\begin{eqnarray}\label{eq:eq3minus}
\frac{\partial \nabla_\perp^2 \zeta_{3-}}{\partial \phi_+} & = \frac{ z_+^2 z_- k_\perp^3}{64 \omega_0 v_A} &\left\{ 
6 \cos[ 2 k_\perp x +  k_\perp y  - k_\parallel \phi_+] 
- 6 \cos[2 k_\perp x + k_\perp y - 2k_\parallel \phi_+ +  k_\parallel \phi_-]  \right.\nonumber\\
& &  +6 \cos[ -2 k_\perp x +  k_\perp y + 2 k_\parallel \phi_+ +  k_\parallel \phi_-] \nonumber\\
 &&-6 \cos[ -2 k_\perp x +  k_\perp y + k_\parallel \phi_+ +2  k_\parallel \phi_-] \\
&&  \left. + 2  \cos[ k_\perp y + k_\parallel \phi_+] 
- 2 \cos[ k_\perp y - k_\parallel \phi_+ +2  k_\parallel \phi_-] 
 \right\} \nonumber  \\
& {\alert{+}} \frac{ z_+ z_-^2 k_\perp^3}{64 \omega_0 v_A}  &\left\{ 
  4 \cos[ k_\perp x + 2 k_\perp y + k_\parallel \phi_-]-4 \cos[ k_\perp x + 2 k_\perp y -  k_\parallel \phi_+ + 2 k_\parallel \phi_-]  \right. \nonumber\\
& &  +4 \cos[- k_\perp x + 2 k_\perp y + k_\parallel \phi_+ + 2 k_\parallel \phi_-]\nonumber\\
&&\left.- 4 \cos[- k_\perp x + 2 k_\perp y + 2 k_\parallel \phi_+ +  k_\parallel \phi_-] \right\},\nonumber 
\end{eqnarray}
and solving for $\zeta_{3\pm}$
\begin{eqnarray}\label{eq:zeta3plusphi}
\zeta_{3+} & = \frac{ z_+^2 z_- k_\perp}{320 \omega_0^2} &\left\{ 
\boxed{8 \omega_0 t \cos[2 k_\perp x + k_\perp y - k_\parallel \phi_+]} + 4 \sin[2 k_\perp x + k_\perp y - 2 k_\parallel \phi_+ + k_\parallel \phi_-] \right.\nonumber \\
&&-4 \sin[2 k_\perp x + k_\perp y - k_\parallel \phi_+]  + 2 \sin[-2 k_\perp x + k_\perp y + k_\parallel \phi_+ + 2 k_\parallel \phi_-] \nonumber \\
&&\left. + 2  \sin[-2 k_\perp x + k_\perp y + 3k_\parallel \phi_+]-4 \sin[-2 k_\perp x + k_\perp y + 2 k_\parallel \phi_+ +  k_\parallel \phi_-] \right\}\nonumber \\
& + \frac{ z_+ z_-^2 k_\perp}{320 \omega_0^2} &\left\{ 
 3 \sin[ k_\perp x + 2 k_\perp y - k_\parallel \phi_+ + 2 k_\parallel \phi_-] 
+ 3 \sin[ k_\perp x + 2 k_\perp y + k_\parallel \phi_+] \right. \\
&&- 6 \sin[ k_\perp x + 2 k_\perp y  + k_\parallel \phi_-]+6 \sin[ -k_\perp x + 2 k_\perp y +2 k_\parallel \phi_+ + k_\parallel \phi_-] \nonumber \\
&&- 3 \sin[ -k_\perp x + 2 k_\perp y + 3 k_\parallel \phi_+] 
-3 \sin[ -k_\perp x + 2 k_\perp y + k_\parallel \phi_+ + 2k_\parallel \phi_-] \nonumber \\
& & \left. +10 \sin[ k_\perp x -2  k_\parallel \phi_+ + k_\parallel \phi_-] 
+ 10  \sin[ k_\perp x - k_\parallel \phi_-] 
-20 \sin[ k_\perp x - k_\parallel \phi_+] \right\},\nonumber
\end{eqnarray}
\begin{eqnarray}\label{eq:zeta3minusphi}
\zeta_{3-} &  = \frac{ z_+^2 z_- k_\perp}{320 \omega_0^2} &\left\{ 
6 \sin[ 2 k_\perp x +  k_\perp y - k_\parallel \phi_+] 
- 3 \sin[ 2 k_\perp x +  k_\perp y - 2 k_\parallel \phi_+ + k_\parallel \phi_-] \right.\nonumber \\
&&- 3 \sin[ 2 k_\perp x +  k_\perp y -  k_\parallel \phi_-] 
+ 6 \sin[ -2 k_\perp x +  k_\perp y +k_\parallel \phi_+ +2 k_\parallel \phi_-] \nonumber \\
&&- 3 \sin[ -2 k_\perp x +  k_\perp y + 2 k_\parallel \phi_+ + k_\parallel \phi_-] 
- 3 \sin[ -2 k_\perp x +  k_\perp y +3  k_\parallel \phi_-] \nonumber \\
& & \left. +20 \sin[ k_\perp y + k_\parallel \phi_-]
- 10 \sin[ k_\perp y -  k_\parallel \phi_+ + 2 k_\parallel \phi_-] 
- 10  \sin[ k_\perp y + k_\parallel \phi_+]  \right\} \nonumber \\
& + \frac{ z_+ z_-^2 k_\perp}{320 \omega_0^2} &\left\{ 
\boxed{{\alert{-}}8 \omega_0 t \cos[ k_\perp x + 2 k_\perp y + k_\parallel \phi_-]}  {\alert{-}}4 \sin[ k_\perp x + 2 k_\perp y -  k_\parallel \phi_+ + 2 k_\parallel \phi_-] \right. \\
&&{\alert{+}}4 \sin[ k_\perp x + 2 k_\perp y + k_\parallel \phi_-] 
+ 2 \sin[- k_\perp x + 2 k_\perp y + 3k_\parallel \phi_-] \nonumber \\
&&\left.+ 2 \sin[- k_\perp x + 2 k_\perp y + 2 k_\parallel \phi_+ +  k_\parallel \phi_-]
 - 4 \sin[- k_\perp x + 2 k_\perp y + k_\parallel \phi_+ + 2 k_\parallel \phi_-] 
 \right\}.\nonumber
\end{eqnarray}

Replacing the characteristic variables with $z$ and $t$ yields 
\begin{eqnarray}\label{eq:zeta3plus}
\zeta_{3+} & = \frac{ z_+^2 z_- k_\perp}{320 \omega_0^2} &\left\{ 
\boxed{8 \omega_0 t \cos[2 k_\perp x + k_\perp y- k_\parallel z -\omega_0 t ]}  \right.\nonumber \\
& &  + 4 \sin[2 k_\perp x + k_\perp y - k_\parallel z -3\omega_0 t] 
-4 \sin[2 k_\perp x + k_\perp y  - k_\parallel z -\omega_0 t ] \nonumber \\
& & + 2 \sin[-2 k_\perp x + k_\perp y + 3 k_\parallel z -\omega_0 t ] 
+ 2  \sin[-2 k_\perp x + k_\perp y  +3 k_\parallel z +3\omega_0 t] \nonumber \\
& &\left. -4 \sin[-2 k_\perp x + k_\perp y  +3 k_\parallel z +\omega_0 t] \right\} \\
& + \frac{ z_+ z_-^2 k_\perp}{320 \omega_0^2} &\left\{ 
3 \sin[ k_\perp x + 2 k_\perp y + k_\parallel z -3\omega_0 t ] 
+ 3 \sin[ k_\perp x + 2 k_\perp y  + k_\parallel z +\omega_0 t ] \right. \nonumber \\
& &- 6 \sin[ k_\perp x + 2 k_\perp y   + k_\parallel z -\omega_0 t ] +6 \sin[ -k_\perp x + 2 k_\perp y  +3 k_\parallel z +\omega_0 t ] \nonumber \\
& &- 3 \sin[ -k_\perp x + 2 k_\perp y  +3 k_\parallel z +3\omega_0 t ] 
-3 \sin[ -k_\perp x + 2 k_\perp y  +3 k_\parallel z -\omega_0 t ] \nonumber \\
& & +10 \sin[ k_\perp x - k_\parallel z -3 \omega_0 t ] 
+ 10  \sin[ k_\perp x - k_\parallel z +\omega_0 t ] \nonumber \\
& &\left. -20 \sin[ k_\perp x - k_\parallel z -\omega_0 t ] \right\},\nonumber
\end{eqnarray}
\begin{eqnarray}\label{eq:zeta3minus}
\zeta_{3-} &  = \frac{ z_+^2 z_- k_\perp}{320 \omega_0^2} &\left\{ 
 6 \sin[ 2 k_\perp x +  k_\perp y - k_\parallel z -\omega_0 t] 
- 3 \sin[ 2 k_\perp x +  k_\perp y - k_\parallel z -3\omega_0 t ] \right.\nonumber \\
& &- 3 \sin[ 2 k_\perp x +  k_\perp y  - k_\parallel z +\omega_0 t]  + 6 \sin[ -2 k_\perp x +  k_\perp y +3 k_\parallel z -\omega_0 t ] \nonumber \\
& &- 3 \sin[ -2 k_\perp x +  k_\perp y  +3 k_\parallel z +\omega_0 t] 
- 3 \sin[ -2 k_\perp x +  k_\perp y  +3 k_\parallel z -3\omega_0 t] \nonumber \\
& & +20 \sin[ k_\perp y + k_\parallel z -\omega_0 t]
- 10 \sin[ k_\perp y  + k_\parallel z -3 \omega_0 t] \nonumber \\
& &\left. - 10  \sin[ k_\perp y + k_\parallel z +\omega_0 t ]  \right\}  \\
& + \frac{ z_+ z_-^2 k_\perp}{320 \omega_0^2} &\left\{ 
\boxed{{\alert{-}}8 \omega_0 t \cos[ k_\perp x + 2 k_\perp y + k_\parallel z -\omega_0 t] } 
{\alert{-}}4 \sin[ k_\perp x + 2 k_\perp y + k_\parallel z -3\omega_0 t] \right. \nonumber \\
& &{\alert{+}}4 \sin[ k_\perp x + 2 k_\perp y + k_\parallel z -\omega_0 t] 
+ 2 \sin[- k_\perp x + 2 k_\perp y +3 k_\parallel z +\omega_0 t] \nonumber \\
& &\left.+ 2 \sin[- k_\perp x + 2 k_\perp y +3 k_\parallel z -3\omega_0 t]
 - 4 \sin[- k_\perp x + 2 k_\perp y  +3 k_\parallel z -\omega_0 t ] 
 \right\}\nonumber,
\end{eqnarray}
and finally the third order solutions for $\V{E}_\perp$ and $\VB_\perp$

\begin{eqnarray}\label{eq:O3B}
\frac{ \V{B}_{\perp 3}}{\sqrt{\mathcal{E}_0}}&  =  \frac{ z_+^2 z_- }{640 v_A^2} 
\frac{k_\perp^2}{k_\parallel^2}
&\left\{  \bigg[ 
\boxed{-8 \omega_0 t\sin(2k_\perp x + k_\perp y - k_\parallel z -\omega_0 t )} 
+ 3\cos(2k_\perp x + k_\perp y - k_\parallel z +\omega_0 t ) \right.  \nonumber \\
&& -10\cos(2k_\perp x + k_\perp y - k_\parallel z -\omega_0 t ) \nonumber \\
& &+7 \cos(2k_\perp x + k_\perp y - k_\parallel z -3\omega_0 t ) \bigg](-\xhat+ 2\yhat)  \nonumber\\
&& \left. 
+ \left[ -2\cos(-2k_\perp x + k_\perp y +3 k_\parallel z +3\omega_0 t ) 
+ \cos(-2k_\perp x + k_\perp y +3 k_\parallel z +\omega_0 t ) \right. \right. \nonumber\\
&&  +4\cos(-2k_\perp x + k_\perp y +3 k_\parallel z -\omega_0 t ) \nonumber \\
& &\left. -3 \cos(-2k_\perp x + k_\perp y +3 k_\parallel z -3\omega_0 t ) \right](\xhat+ 2\yhat)  \nonumber\\
&& 
+ \left[-10\cos(k_\perp y + k_\parallel z +\omega_0 t ) 
+20 \cos(k_\perp y + k_\parallel z -\omega_0 t ) \right. \nonumber \\
&&\left. \left. - 10 \cos(k_\perp y + k_\parallel z -3 \omega_0 t )\right]\xhat \right\}  \\
&+ \frac{ z_+ z_-^2 }{640 v_A^3} 
\frac{k_\perp^2}{k_\parallel^2} &\left\{ \bigg[ 
\boxed{{\alert{-}}8 \omega_0 t\sin(k_\perp x + 2 k_\perp y + k_\parallel z -\omega_0 t )}
+ 3\cos(k_\perp x + 2 k_\perp y + k_\parallel z +\omega_0 t )  \right. \nonumber\\
&&  {\alert{-10}}\cos(k_\perp x + 2 k_\perp y + k_\parallel z -\omega_0 t ) \nonumber \\
&&{\alert{+7}} \cos(k_\perp x + 2 k_\perp y + k_\parallel z -3\omega_0 t ) \bigg](-2\xhat+ \yhat)   \nonumber\\
&& \left. 
+ \left[ 3\cos(-k_\perp x + 2 k_\perp y +3 k_\parallel z +3\omega_0 t ) 
-4 \cos(-k_\perp x + 2k_\perp y +3 k_\parallel z +\omega_0 t ) \right. \right. \nonumber\\
&&  -\cos(-k_\perp x +2 k_\perp y +3 k_\parallel z -\omega_0 t )\nonumber\\
&& \left. +2 \cos(-k_\perp x + 2k_\perp y +3 k_\parallel z -3\omega_0 t ) \right](2\xhat+ \yhat)  \nonumber\\
&& 
+ \left[ 10\cos(k_\perp x - k_\parallel z +\omega_0 t ) 
-20 \cos(k_\perp x - k_\parallel z -\omega_0 t ) \right. \nonumber\\
&& \left. \left. + 10 \cos(k_\perp x - k_\parallel z -3 \omega_0 t )\right]\yhat \right\}, \nonumber
\end{eqnarray}

\begin{eqnarray}\label{eq:O3E}
\frac{ c \V{E}_{\perp 3}}{B_0} & =   \frac{ z_+^2 z_- }{640 v_A^2} 
\frac{k_\perp^2}{k_\parallel^2}
&\left\{ \bigg[ 
\boxed{8 \omega_0 t\sin(2k_\perp x + k_\perp y - k_\parallel z -\omega_0 t )}
+ 3\cos(2k_\perp x + k_\perp y - k_\parallel z +\omega_0 t )  \right.\nonumber\\
&&  -2\cos(2k_\perp x + k_\perp y - k_\parallel z -\omega_0 t )\nonumber\\
&& -\cos(2k_\perp x + k_\perp y - k_\parallel z -3\omega_0 t ) \bigg](2\xhat+ \yhat)  \nonumber\\
&& \left. 
+ \left[ -2\cos(-2k_\perp x + k_\perp y +3 k_\parallel z +3\omega_0 t ) 
+ 7\cos(-2k_\perp x + k_\perp y +3 k_\parallel z +\omega_0 t ) \right. \right. \nonumber\\
&&  -8\cos(-2k_\perp x + k_\perp y +3 k_\parallel z -\omega_0 t ) \nonumber\\
&&\left.+3 \cos(-2k_\perp x + k_\perp y +3 k_\parallel z -3\omega_0 t ) \right](-2\xhat+ \yhat)   \nonumber\\
&& 
+ \left[10\cos(k_\perp y + k_\parallel z +\omega_0 t ) 
-20 \cos(k_\perp y + k_\parallel z -\omega_0 t ) \right. \nonumber\\
&&\left. \left.+ 10 \cos(k_\perp y + k_\parallel z -3 \omega_0 t )\right]\yhat \right\}  \\
&  + \frac{ z_+ z_-^2 }{640 v_A^3} 
\frac{k_\perp^2}{k_\parallel^2}  & \left\{ \bigg[
\boxed{{\alert{-}}8 \omega_0 t\sin(k_\perp x + 2k_\perp y + k_\parallel z -\omega_0 t )}
- 3\cos(k_\perp x + 2k_\perp y + k_\parallel z +\omega_0 t ) \right.  \nonumber\\
&&  {\alert{+2}}\cos(k_\perp x + 2k_\perp y + k_\parallel z -\omega_0 t )\nonumber\\
&&{\alert{+}} \cos(k_\perp x + 2k_\perp y + k_\parallel z -3\omega_0 t ) \bigg] (\xhat+2\yhat)  \nonumber\\
&& \left. 
+ \left[ 3\cos(-k_\perp x + 2k_\perp y +3 k_\parallel z +3\omega_0 t ) 
-8\cos(-k_\perp x + 2k_\perp y +3 k_\parallel z +\omega_0 t ) \right. \right. \nonumber\\
&&  +7\cos(-k_\perp x + 2k_\perp y +3 k_\parallel z -\omega_0 t )\nonumber\\
&& \left.-2 \cos(-k_\perp x + 2k_\perp y +3 k_\parallel z -3\omega_0 t ) \right](-\xhat+ 2\yhat)   \nonumber\\
&&  
+ \left[ -10\cos(k_\perp x - k_\parallel z +\omega_0 t ) 
+20 \cos(k_\perp x - k_\parallel z -\omega_0 t ) \right.\nonumber\\
&&\left. \left. -10 \cos(k_\perp x - k_\parallel z -3 \omega_0 t )\right]\xhat  \right\}.\nonumber
\end{eqnarray}


\end{document}